\documentclass[twoside,journal]{IEEEtran}
\usepackage{amsfonts}
\usepackage{amssymb}
\usepackage{amsmath}
\usepackage{amsthm}
\usepackage{mathrsfs}
\usepackage{latexsym}
\usepackage{verbatim}
\usepackage{subfigure}
\usepackage{balance}
\usepackage{booktabs}
\usepackage{fancybox}
\usepackage{bm}
\usepackage{balance}
\usepackage{changepage}
\usepackage{extarrows}
\usepackage{enumerate}
\usepackage{algorithm}
\usepackage{algpseudocode}
\usepackage{multirow}
\usepackage{array}
\usepackage{graphicx}
\usepackage{epsfig}
\usepackage{epstopdf}
\usepackage{caption}
\usepackage{color,soul}
\usepackage{verbatim}
\usepackage{cite}
\usepackage[numbers,sort&compress]{natbib}
\newcommand{\bh}{{\bf h}}
\newcommand{\bQ}{{\bf Q}}
\newcommand{\bH}{{\bf H}}
\newcommand{\bw}{{\bf w}}
\newcommand{\bq}{{\bf q}}
\newcommand{\bW}{{\bf W}}
\newcommand{\bM}{{\bf M}}
\newcommand{\bA}{{\bf A}}
\newcommand{\bB}{{\bf B}}
\newcommand{\bF}{{\bf F}}
\newcommand{\bD}{{\bf D}}
\newcommand{\bd}{{\bf d}}
\newcommand{\bC}{{\bf C}}
\newcommand{\bc}{{\bf c}}
\newcommand{\bG}{{\bf G}}
\newcommand{\bU}{{\bf U}}
\newcommand{\bu}{{\bf u}}
\newcommand{\bR}{{\bf R}}
\newcommand{\bv}{{\bf v}}
\newcommand{\bx}{{\bf x}}
\newcommand{\by}{{\bf y}}
\newcommand{\bs}{{\bf s}}
\newcommand{\bI}{{\bf I}}
\newcommand{\bn}{{\bf n}}
\newcommand{\ba}{{\bf a}}
\newcommand{\bX}{{\bf X}}
\newcommand{\bZ}{{\bf Z}}
\newcommand{\bE}{{\bf E}}
\newcommand{\bK}{{\bf K}}
\newcommand{\br}{{\bf r}}
\newcommand{\bt}{{\bf t}}
\newcommand{\bS}{{\bf S}}
\newcommand{\bz}{{\bf z}}
\newcommand{\bg}{{\bf g}}
\newcommand{\bJ}{{\bf J}}

\newcommand{\be}{{\bf e}}
\newcommand{\bp}{{\bf p}}
\newcommand{\bY}{{\bf Y}}
\begin{document}
	\title{Reconfigurable Intelligent Surface-Aided Secure Integrated Radar and Communication Systems}
		\author{Tong-Xing Zheng, \emph{Member, IEEE}, Xin Chen, Lan Lan, \emph{Member, IEEE}, Ying Ju, \emph{Member, IEEE}, Xiaoyan Hu, \emph{Member, IEEE}, Rongke Liu, \emph{Senior Member, IEEE}, Derrick Wing Kwan Ng, \emph{Fellow, IEEE}, and Tiejun Cui, \emph{Fellow, IEEE}
			
		\thanks{The work of T.-X. Zheng was supported by the Aeronautical Science Foundation of China under Grant ASFC-2022Z021070001. The work of L. Lan was supported in part by the National Natural Science Foundation of China under Grant 62471348. The work of Y. Ju was supported by the National Natural Science Foundation of China under Grant 62102301. The work of X. Hu was supported in part by the National Natural Science Foundation of China under Grants 62201449 and 62471380, in part by the Key R\&D Projects of Shaanxi Province under Grant 2023-YBGY-040. The work of R. Liu was supported in part by Shenzhen Fundamental Research Project under Grant JCYJ20220818103413029, in part by Guangdong Basic and Applied Basic Research Foundation 2023B1515120093 and in part by Shenzhen Science and Technology R\&D Funds under Grant JSGG20220831100602005. The work of T. Cui was supported by the National Natural Science Foundation of China under Grant 62288101. Part of this work has been presented at the IEEE ICCWorkshops 2022 \cite{9814654} and IEEE ICCC 2023 \cite{10233430}. The associate editor coordinating the review of this paper and approving it for publication was Dr. Guido Carlo Ferrante. (\emph{Corresponding authors: Tong-Xing Zheng; Xin Chen}.)}
		\thanks{
		T.-X. Zheng and X. Hu are with the School of Information and Communications Engineering, Xi'an Jiaotong University, Xi'an 710049, China. T.-X. Zheng is also with the Ministry of Education Key Laboratory for Intelligent Networks and Network Security, Xi’an Jiaotong University, Xi’an 710049, China (e-mail: zhengtx@mail.xjtu.edu.cn; xiaoyanhu@xjtu.edu.cn).}
		\thanks{
		X. Chen is with the Purple Mountain Laboratories, Nanjing 211111, China (e-mail: chenxin02@pmlabs.com.cn).}
		\thanks{L. Lan is with the National Key Laboratory of Radar Signal Processing, Xidian University, Xi’an 710071, China (email: lanlan@xidian.edu.cn).}
		\thanks{Y. Ju is with the School of Telecommunications Engineering, Xidian University, Xi’an 710071, China (email: juying@xidian.edu.cn).}
		\thanks{R. Liu is with the School of Electronic and Information Engineering, Beihang University, Beijing 100191, China (e-mail: rongke$\_$liu@buaa.edu.cn).} 
		\thanks{D. W. K. Ng is with the School of Electrical Engineering and Telecommunications, University of New South Wales, Sydney, NSW 2052, Australia (e-mail: w.k.ng@unsw.edu.au).}
		\thanks{T. Cui is with the School of Information Science and Engineering, Southeast University, Nanjing 210096, China (e-mail: tjcui@seu.edu.cn).}
		
	}
	\maketitle
	
	\begin{abstract}
		Despite the enhanced spectral efficiency brought by the integrated radar and communication technique, it poses significant risks to communication security when confronted with malicious radar targets. To address this issue, a reconfigurable intelligent surface (RIS)-aided transmission scheme is proposed to improve secure communication in two systems, i.e., the radar and communication co-existing (RCCE) system, where a single transmitter is utilized for both radar sensing and communication, and the dual-functional radar and communication (DFRC) system. At the design stage, optimization problems are formulated to maximize the secrecy rate while satisfying the radar detection constraint via joint active beamforming at the base station and passive beamforming of RIS in both systems. Particularly, a zero-forcing-based block coordinate descent (BCD) algorithm is developed for the RCCE system. Besides, the Dinkelbach method combined with semidefinite relaxation is employed for the DFRC system, and to further reduce the computational complexity, a Riemannian conjugate gradient-based alternating optimization algorithm is proposed. Moreover, the RIS-aided robust secure communication in the DFRC system is investigated by considering the eavesdropper's imperfect channel state information (CSI), where a bounded uncertainty model is adopted to capture the angle error and fading channel error of the eavesdropper, and a tractable bound for their joint uncertainty is derived. Simulation results confirm the effectiveness of the developed RIS-aided transmission scheme to improve the secrecy rate even with the eavesdropper's imperfect CSI, and comparisons between both systems reveal that the RCCE system can provide a higher secrecy rate than the DFRC system.  
	\end{abstract}
	
	\section{Introduction}
	Recently, the integrated radar and communication technique has garnered significant attention as a valuable technique to address the escalating challenge of spectrum scarcity and achieve mutual benefits in communication and sensing \cite{8288677}. {{To unlock its potential, two practical systems that accommodate both radar sensing and communication functions have been proposed, i.e., the radar and communication co-existing (RCCE) systems \cite{6263762} and dual-funcational radar and communication (DFRC) systems \cite{7347464}. Specifically, the RCCE system transmits two different signals occupied the same time and/or frequency resources by a transmitter, one for radar sensing and the other for communication. It focuses on developing efficient interference management techniques to ensure that radar sensing and communication can operate smoothly without interfering with each other. Various techniques such as opportunistic spectrum sharing \cite{6263762} and null-space projection method \cite{6503914} are proposed to support the co-existence of radar and communication. Differently, DFRC system transmits a single signal for both sensing and communication, offering a more favorable approach to integrate sensing and communication, which can further improve spectral efficiency and reduce hardware costs.}}
	
	{{Despite the improved spectral efficiency and lower hardware costs, wireless communication security is a troubling problem in the integrated radar and communication system. Due to the broadcast nature of wireless channels, wireless communication is inherently vulnerable to eavesdropping attacks. Compared with radar and communication co-existence system, DFRC system faces greater eavesdropping risks. As we know, a high-power main beam is desired for target detection. To achieve high-quality sensing, the transmitter must design a signal to synthesize such a high-power beam towards targets. In DFRC system, a single transmitted signal is employed for both radar sensing and communication. If targets are malicious eavesdroppers, the communication information embedded in the signal will be highly likely to be wiretapped. Therefore, striking an optimal balance between secure communication and radar sensing is a crucial task when designing integrated radar and communication systems.}} 
	
	\subsection{Related Work}
	The current state-of-the-art works of PLS for integrated radar and communication, RIS-aided PLS, and RIS-aided DFRC are reviewed as follows.
	\subsubsection{PLS for integrated radar and communication}
	Physical layer security (PLS) is an innovative technique to ensure secure communication \cite{book1}. Recently, PLS has been introduced into integrated radar and communication systems to guarantee secure communication \cite{8327462,CHALISE2018282,9538912,9199556,9737364}. Specifically, secure communication is investigated in a DFRC system, where both the information-embedded signal desired for a legitimate user and the signal embedded with false information to confuse an eavesdropping radar target are transmitted by a multi-input multi-output (MIMO) radar for target detection \cite{8327462}. In this respect, problems in terms of secrecy rate maximization, radar received signal-to-interference-plus-noise ratio (SINR) maximization, and transmit power minimization are investigated. In \cite{CHALISE2018282}, a unified system of communications and passive radar is considered, where radar and communication signals are transmitted via a joint transmitter with different resource scheduling schemes. Specifically, the SINR at radar receiver is maximized while the secrecy rate is ensured above a certain threshold \cite{CHALISE2018282}. Furthermore, in \cite{9538912,9199556}, artificial noise (AN) is utilized to assist secure communication for a DFRC system and both imperfect channel state information (CSI) and target location uncertainty are taken into account \cite{9199556}. Besides, the multi-user interference is exploited to enhance secure communication in DFRC systems, where the constructive interference is utilized to enhance the received signals at communication users and destructive interference is adopted to deteriorate the eavesdropped signals at radar target \cite{9737364}.   
	
	\subsubsection{RIS-PLS}
	{{Although the aforementioned PLS techniques such as AN and multi-antenna can realize secure communication, they are generally channel-dependent. In other words, their performance is heavily influenced by wireless channels. Additionally, whether transmitting AN or increasing the number of transmit antenna will lead to higher power consumption. 
	Reconfigurable intelligent surface (RIS) is a software-controlled meta-surface composed of numerous passive reflecting elements. Each reflecting element is able to independently adjust the phase and/or amplitude of incident signals, thus realizing smart reconfiguration of wireless channels. Besides, RIS reflects the signals passively with low power consumption. Actually, RIS-aided PLS has gained significant research attention \cite{8723525,8742603,9159923}. Deploying RIS in wireless systems, by designing the reflecting coefficients of RIS, the reflected signal and direct signal can be constructively added up at the desired users, and destructively combined at the eavesdroppers, which provides a new approach to enhance secure communication.}}
	
	\subsubsection{RIS-DFRC}
	 On the other hand, RIS has been employed in DFRC systems. Specifically, an RIS-aided DFRC system is examined in \cite{9743509} by jointly designing the transmit beamforming and RIS phase shift matrix to maximize the signal-to-noise ratio (SNR) at the radar receiver while considering the transmit power and communication rate constraints. In \cite{9769997}, a RIS-aided DFRC system is considered, which enables concurrent multi-user communication and MIMO radar detection. Moreover, due to the high computational complexity of traditional optimization algorithms, such as space time adaptive processing (STAP) and alternative direction method of multipliers (ADMM), a deep reinforcement learning-based algorithm is proposed in \cite{9733335} to optimize the transmit beamforming and RIS phase shifts, aiming to maximize the capacity of an RIS-aided DFRC system with multiple users in the THz band. 
	 
	 {{To the best of our knowledge, few papers have investigated RIS-aided secure communication in integrated radar and communication system. There is lack of unified design framework and efficient parameter optimization algorithms for it. As mentioned earlier, the integrated radar and communication system are mainly divided into two categories: RCCE system and DFRC system, for which tailored secure communication schemes need to be designed. Comparison of secrecy performance between them is also worth exploring. Therefore, we carry out research on RIS-aided secure communication in integrated radar and communication system. It is worth noting that although some papers have studied RIS-assisted secure communication in integrated radar and communication system \cite{9685487,9747551}, the obtained results are based on target’s perfect CSI which is not practical. It is difficult for the BS to get perfect CSI and accurate location of targets in practice. Since the RIS-assisted secure communication is particularly sensitive to the channel characteristics difference between users and eavesdroppers, the robust RIS-aided secure communication scheme under imperfect CSI is also studied in the paper.}}
	  
	\subsection{Our Work and Contributions}
	Based on the aforementioned motivations, the RIS-aided secure communication in integrated radar and communication systems is investigated. In particular, two systems are considered: the RCCE system and DFRC system. To shed light on the design, we first assume that the target's perfect CSI is available. For both systems, the secrecy rate is supposed to be maximized by jointly optimizing the transmit beamforming of the BS and the phase shifts of RIS while ensuring radar detection constraint. Moreover, considering that BS may not be able to acquire accurate CSI of the target in practice, the RIS-aided robust secure communication in the integrated radar and communication system with imperfect target’s CSI is also investigated. 
	% It is worth noting that this design framework can be extended to radar and communication co-existing systems.
	The main contributions of our paper are summarized as follows.
	
	\begin{itemize}
	\item The RIS-aided secure communication in the RCCE system and DFRC system are investigated, where the corresponding beamforming design schemes to maximize the secrecy rate are studied. {{A low-complexity RCG-based AO algorithm is developed to handle the formulated problem for DFRC system.}} Secrecy performance comparison between both systems are conducted in simulations.
	
	\item {{The RIS-aided robust secure communication in the integrated radar and communication system is investigated by taking into account the target’s imperfect CSI. A tractable bound for the joint uncertainty of angle and fading channel estimation errors of target is derived that facilitates the development of a BCD-based algorithm combined with S-procedure to obtain a suboptimal solution to the secrecy rate maximization problem.}}
	
	\item Simulation results show that RIS can help achieve higher secrecy rate. Compared to DFRC system, the RCCE system performs better in terms of secrecy rate. Moreover, it is observed that the secrecy rate significantly decreases with target’s imperfect CSI, highlighting the importance of considering CSI uncertainty for system design.    
	\end{itemize} 

	\subsection{Organization and Notation}
	The remainder of this paper is organized as follows: Sections II and III formulate the secrecy rate maximization problems and provide corresponding solutions for RCCE and DFRC systems, respectively. Section IV investigates the RIS-aided robust secure communication for DFRC system with target's imperfect CSI. Simulation results are provided in Section V, and conclusions are drawn in Section VI.   
	
	\emph{Notations}: Unless further specified, bold uppercase (lowercase) letters denote the matrices (vectors); $\mathbb{C}$ and $\mathbb{E}$ denote the set of complex numbers and the expectation operation, respectively; $\mathcal{CN}\left(\bm{\mu},\bm{\Sigma}\right)$ denotes the complex Gaussian distribution with mean $\bm{\mu}$ and covariance $\bm{\Sigma}$; $\left(\cdot\right)^H$ and $\left(\cdot\right)^T$ are the conjugate transpose and transpose, respectively; $\mathrm{diag}\left(\bx\right)$ denotes a diagonal matrix composed of the elements in vector $\bx$; $\|\cdot\|$ and $\left|\cdot\right|$ stand for the $l_2$ norm of a vector and the modulus of a scalar, respectively; $\left[\cdot\right]^+$ represents $\mathrm{max}\left(0, \cdot\right)$; $\mathrm{tr}\left(\cdot\right)$ and $\mathrm{rank}\left(\cdot\right)$ denote the trace and the rank of a matrix, respectively; $\mathfrak{R}\left(\cdot\right)$ denotes the real part of the argument; $ \otimes$ and $\odot$ denote the Kronecker and Hadamard product respectively; $\mathrm{vec}\left(\bX\right)$ is the vectorization of matrix $\bX$; $\mathrm{Diag}\left(\bX\right)$ denotes a vector composed of the diagonal elements of matrix $\bX$; $\left(\cdot\right)^*$ denotes the conjugate; $\Vert\cdot\Vert_F$ denotes the Fibonacci norm.

\section{Secure Communication in RCCE System}
\begin{figure}[htbp]
	\centerline{\includegraphics[scale = 0.22]{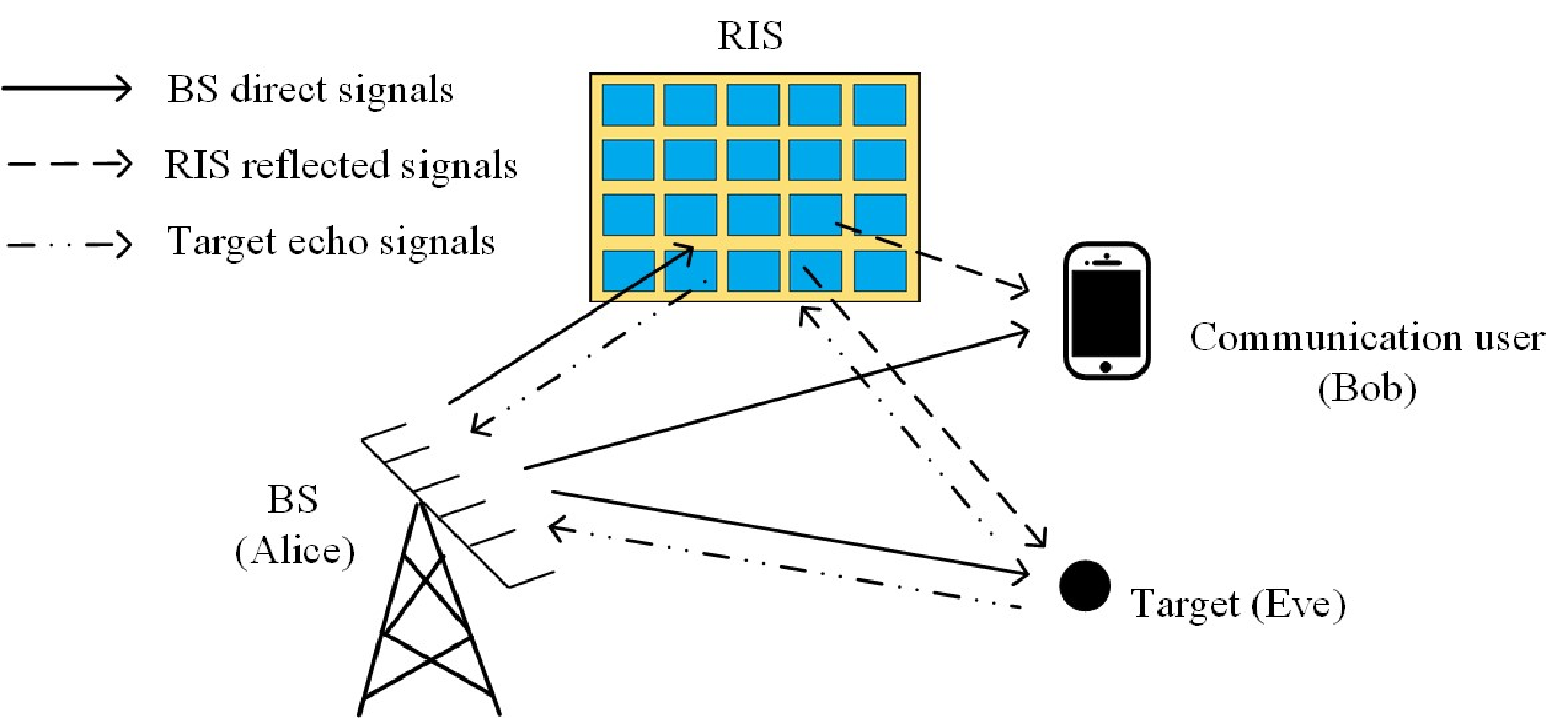}}
	\caption{An illustration of a RIS-aided secure integrated radar and communication system.} \label{fig1}
\end{figure}

We consider a RIS-aided secure integrated radar and communication system, as depicted in Fig. \ref{fig1}. {{A BS, Alice, equipped with $N$ antennas serves a single-antenna communication user, Bob, as well as detecting a single-antenna target, Eve. We assume that Eve might be a potential eavesdropper, who tries to intercept the information transmitted from Alice to Bob.}} A RIS comprising $M$ elements is deployed to safeguard against eavesdropping. Based on the system model, we investigate the RIS-aided secure communication in the settings of RCCE system and DFRC system, respectively. 
\captionsetup[figure]{name={Fig.},labelsep=period}

We first examine the RIS-aided secure transmission in RCCE system, where the BS simultaneously transmits separate radar and communication signals utilizing the same spectral resources to realize concurrent radar detection and communication. A secrecy rate maximization problem is formulated and a ZF-based BCD algorithm is proposed to address the problem at hand.    

\subsection{Problem Formulation}
 In RCCE system, the received signals at Bob and Eve can be expressed as
\begin{equation}
	y_\mathrm{B} = \sqrt{P}\left(\bh_{\mathrm{IB}}^H\bQ\bH_{\mathrm{AI}} + \bh_{\mathrm{AB}}^H \right)\bx + n_\mathrm{B},
\end{equation} 
\begin{equation}
	y_\mathrm{E} = \sqrt{P}\left(\bh_{\mathrm{IE}}^H\bQ\bH_{\mathrm{AI}} + \bh_{\mathrm{AE}}^H \right)\bx + n_\mathrm{E},
\end{equation} 
respectively, where $P$ denotes the total transmit power from the BS; $\bx = \left(\sqrt{\epsilon}\bw b + \sqrt{1 - \epsilon}\bs\right)$;  
$b \in \mathbb{C}, \mathbb{E}\left[bb^H\right] = 1$ is the transmitted communication signal;
%$\mathbb{E}\left[bb^H\right] = 1$; 
$\bw\in\mathbb{C}^{N\times 1}$ represents the beamforming vector for communication signal at the BS with $\|\bw\| = 1$; $\bs\in\mathbb{C}^{N\times 1}$ represents the radar signal; $\epsilon \in \left[0,1\right]$ denotes the proportion of the power allocated to communication; $\bH_{\mathrm{AI}}\in\mathbb{C}^{M\times N}$, $\bh_{\mathrm{AB}} \in\mathbb{C}^{N\times 1}$, $\bh_{\mathrm{AE}}\in\mathbb{C}^{N\times 1}$, $\bh_{\mathrm{IB}}\in\mathbb{C}^{M\times 1}$, $\bh_{\mathrm{IE}}\in\mathbb{C}^{M\times 1}$ denote the channels of Alice-RIS, Alice-Bob, Alice-Eve, RIS-Bob, and RIS-Eve, respectively; $\bQ  \!\!=\!\! \mathrm{diag}(\bq) \!\!=\!\! \mathrm{diag}\left(\left[e^{j\theta_1},e^{j\theta_2},\dots,e^{j\theta_M}\right]^T\right)$ denotes the phase shift matrix of RIS with $\theta_i \in \left[0,2\pi\right)$ being the phase shift of the $i$-th reflecting element, $i \in \{1,\dots, M\}$; $n_\mathrm{B} \!\sim \!\mathcal{CN}\left(0,\sigma_\mathrm{B}^2\right)$ and $n_\mathrm{E} \!\sim \!\mathcal{CN}\left(0,\sigma_\mathrm{E}^2\right)$ denote the received noise at Bob and Eve, respectively. {{Assuming that Eve is located at an open area where the radio propagation environment includes almost no scatterers, the channel $\bh_{\mathrm{AE}}$ is modeled as a line of sight (LoS) channel following the traditional setting in the radar field as done in \cite{9199556,9769997,9743509}}} which is given by
\begin{equation}
	\label{a_A}
	\bh_{\mathrm{AE}} = \beta_{\mathrm{AE}}\ba_{\mathrm{A}},
\end{equation}
where $\beta_{\mathrm{AE}}$ denotes the path loss and $\ba_{\mathrm{A}}$ denotes the transmit array steering vector which is expressed as 
\begin{equation}
	\ba_{\mathrm{A}} = \left[1,e^{j\frac{2\pi d}{\lambda}\sin\theta},\dots,e^{j\frac{2\pi d}{\lambda}\left(N - 1\right)\sin\theta}\right]^T,
\end{equation}
where $\lambda$ is the wavelength; $d$ is the spacing between two adjacent elements of transmit array; $\theta$ specifies the direction of potential eavesdropper\footnote{The location of the target can be obtained by prior radar sensing. As done in \cite{9737357, 9933849}, the BS adopts a multi-stage sensing scheme to obtain the exact positions of targets.}. Assuming Bob is located in a hot-spot area and RIS is deployed around to assist secure communication, the Rayleigh fading channels are adopted to model other channels \cite{9199556,9685487}.\footnote{{We assume that the channels are quasi-static and block-fading, i.e., the channels vary from frame to frame, but remain unchanged within a frame. In the paper, we focus on the parameter design within a frame and the channels in a frame are deterministic and known by channel estimation.}} For example, $\bh_{\mathrm{AB}} = \beta_{\mathrm{AB}}\bg_{\mathrm{AB}}$ where $\beta_{\mathrm{AB}}$ denotes the path loss and $\bg_{\mathrm{AB}}\in \mathbb{C}^{N\times1} \!\sim\! \mathcal{CN}\left(\bf{0},\bI\right)$ denotes the Rayleigh fading. The communication signals are statistically independent of radar signals. {{Moreover, we first assume that the perfect CSI is available at the BS}}\footnote{{In practice, it is generally difficult to obtain the perfect CSI between BS/RIS and the target. The results based on perfect CSI serve as the theoretical performance upper bounds for the considered system, which provide guidance for following robust design under imperfect CSI.}}. 

Based on the received signals, the achievable communication rate at Bob and Eve can be expressed as
\begin{equation}
	\begin{aligned}
	C_{\mathrm{J}} & = \log_2\left(1 + \frac{P_{\mathrm{cJ}}}{P_{\mathrm{rJ}} + \sigma_\mathrm{J}^2}\right) \\
	& = \log_2\left(1 + \frac{\epsilon P\bh_\mathrm{J}\bw \bw^H\bh_\mathrm{J}^H}{\left(1 - \epsilon\right)P\bh_{\mathrm{J}}\bR\bh_{\mathrm{J}}^H +\sigma_\mathrm{J}^2}\right), \mathrm{J}\in\{\mathrm{B},\mathrm{E}\},
	\end{aligned}
\end{equation} 
%\begin{equation}
%	\begin{aligned}
%		C_{\mathrm{E}} & = \log_2\left(1 + \frac{P_{\mathrm{cE}}}{P_{\mathrm{rE}} + \sigma_\mathrm{E}^2}\right) \\
%		& = \log_2\left(1 + \frac{\epsilon P\bh_{\mathrm{E}}\bw \bw^H\bh_{\mathrm{E}}^H}{\left(1 - \epsilon\right)P\bh_{\mathrm{E}}\bR\bh_{\mathrm{E}}^H +\sigma_\mathrm{E}^2}\right),
%	\end{aligned}
%\end{equation}
where $P_{\mathrm{cB}}$/$P_{\mathrm{cE}}$ represents the received communication signal power at Bob/Eve; $P_{\mathrm{rB}}$/$P_{\mathrm{rE}}$ represents the received radar signal power at Bob/Eve; $\bh_{\mathrm{B}} \!=\! \bh_{\mathrm{IB}}^H\bQ\bH_{\mathrm{AI}} + \bh_{\mathrm{AB}}^H$ and $\bh_{\mathrm{E}} \!=\! \bh_{\mathrm{IE}}^H\bQ \bH_{\mathrm{AI}} + \bh_{\mathrm{AE}}^H$; $\bR = \mathbb{E}\left[\bs\bs^H\right]$ denotes the radar transmit covariance matrix. {{Since the target is a potential eavesdropper, the secrecy rate is defined by the difference between $C_{\mathrm{B}}$ and $C_{\mathrm{E}}$ which is expressed as $C_s = \left[C_{\mathrm{B}} - C_{\mathrm{E}}\right]^+$ \cite{9199556}.}} 

On the other hand, for the radar detection, the echo signals received at Alice can be expressed as
\begin{equation}
	\by_\mathrm{A} = \sqrt{P}\bh_{\mathrm{E}}^H\bh_{\mathrm{E}}\bx + \bn_{\mathrm{A}},
\end{equation}
where $\bn_{\mathrm{A}} \sim \mathcal{CN}\left({\bf 0}, \sigma_{\mathrm{A}}^2 \bI\right)$ denotes the received noise at Alice. We ignore the echo signals reflected by RIS more than once since they are severely attenuated due to the associated path loss. Hence, the received SINR at Alice can be calculated as
\begin{equation}
	\gamma_\mathrm{A} = \frac{P_{\mathrm{rA}}}{P_{\mathrm{cA}} + N\sigma_{\mathrm{A}}^2} = \frac{\left(1 - \epsilon\right)P\mathrm{tr}\left(\bH_{\mathrm{E}}^H\bH_{\mathrm{E}}\bR\right)}{\epsilon P\bw^H\bH_{\mathrm{E}}^H\bH_{\mathrm{E}}\bw + N\sigma_{\mathrm{A}}^2},
\end{equation}
where $P_{\mathrm{rA}}$ and $P_{\mathrm{cA}}$ denote the received radar and communication power at Alice, respectively, and $\bH_{\mathrm{E}} \!=\! \bh_{\mathrm{E}}^H\bh_{\mathrm{E}}$. To realize reliable radar detection, the received SINR at Alice should be no less than a predetermined threshold $\gamma$, i.e., $\gamma_{\mathrm{A}} \geq \gamma$.

Focusing on secure communication, we aim to maximize the secrecy rate while maintaining the radar detection performance by jointly optimizing $\epsilon$, $\bw$, $\bR$, and $\bQ$ that yields the following problem formulation:
\begin{subequations}
	\begin{align}	\label{p_}
		\mathcal{P}_1:
		\max_{\epsilon,\bw,\bR,\bQ}~
		&C_s\left(\epsilon,\bw,\bR,\bQ\right) = C_{\mathrm{B}} - C_{\mathrm{E}}\\
		~\mathrm{s.t.}
		\label{p_c_ep}
		~~& \epsilon\in\left[0,1\right],\\
		\label{p_c1}
		~~& \|\bw\| = 1,\\
		\label{p_c4}
		~~& \bR = \bR^H, \bR \succeq {\bf 0}, \mathrm{tr}\left(\bR\right) = 1,\\
		\label{p_c2}
		~~& \left|\bQ_{i,i} \right| = 1, \forall i=1,2,\dots,M,\\
		\label{p_c3}
		~~& \gamma_{\mathrm{A}}  \geq \gamma,
	\end{align}
\end{subequations}
where \eqref{p_c1} denotes the unit norm constraint of $\bw$, \eqref{p_c4} denotes the radar transmit power constraint, \eqref{p_c2} specifies the unit modulus constraint of $\bQ_{i,i}$, and \eqref{p_c3} represents the radar detection constraint. Note that we have dropped the operator $\left[\cdot\right]^+$ in the objective function without loss of optimality \cite{9199556,8723525}.  

\subsection{Problem Solution}
{{Due to the non-convex objective function, non-convex constraints \eqref{p_c2}, and \eqref{p_c3} where multiple optimization variables are non-trivially coupled, problem $\mathcal{P}_1$ is a non-convex problem which is hard to be solved optimally.}} To address this challenge, a suboptimal ZF-based BCD algorithm is proposed. 

%In the inner layer, we fix $\epsilon$ and a BCD algorithm is applied to solve the following problem
%\begin{subequations}
%	\begin{align}	
%		\mathcal{P}_1:
%		\max_{\bw,\bR,\bQ}~
%		&R_s\left(\bw,\bR,\bQ\right) = R_B - R_E\\
%		~\mathrm{s.t.}
%		~~&\eqref{p_c1}-\eqref{p_c3},
%	\end{align}
%\end{subequations}
Specifically, the entire optimization variables in problem $\mathcal{P}_1$ are partitioned into four blocks, i.e., $\bw$, $\bR$, $\bQ$, and $\epsilon$, and we maximize $C_s$ by alternatingly optimizing each block with the other three blocks fixed until the objective function in $\mathcal{P}_1$ converges. Next, we introduce the BCD algorithm in details.

\subsubsection{Optimizing $\bw$}
For simplicity, a ZF beamforming is adopted at Alice and no communication signals are leaked to Eve. The ZF beamforming vector $\bw$ is given by 
\begin{equation} \label{zf_w}
	\bw = \frac{\left(\bI -\frac{\bH_{\mathrm{E}}}{\left\|\bh_{\mathrm{E}}\right\|^2}\right)\bh_{\mathrm{B}}^H}{\left\|\left(\bI -\frac{\bH_{\mathrm{E}}}{\left\|\bh_{\mathrm{E}}\right\|^2}\right)\bh_{\mathrm{B}}^H\right\|}.
\end{equation}

\subsubsection{Optimizing $\bR$}
Motivated by the null-space projection method \cite{6503914}, we force the radar signals, to fall into the nullspace of the effective channel between Alice and Bob, $\bh_{\mathrm{B}}$, so as to eliminate the interference brought by radar to communication, i.e., $\mathrm{tr}\left(\bH_{\mathrm{B}}\bR\right) = 0$ with $\bH_{\mathrm{B}} = \bh_{\mathrm{B}}^H\bh_{\mathrm{B}}$ holds. Thus, given $\epsilon$, $\bw$, and $\bQ$, the problem of optimizing $\bR$ can be expressed as
\begin{subequations}	\label{p_6}
	\begin{align}
	\mathcal{P}_2:
	\max_{\bR}~
	C_s\left(\bR\right) & =  ~C_{\mathrm{B}} - C_{\mathrm{E}}  = f_1 - f_2 -f_3 + f_4 \\
	\notag
	 \mathrm{s.t.} ~& \eqref{p_c4}, \eqref{p_c3}, \\
	\label{cons_R}
	~~~&\mathrm{tr}\left(\bH_{\mathrm{B}}\bR\right) = 0,
	\end{align}		
\end{subequations}
where $f_1 = \log_2\left(x_\mathrm{B} + \epsilon\mathrm{tr}\left(\bH_{\mathrm{B}}\bW\right)\right), f_2 = \log_2\left(x_\mathrm{B}\right), f_3 = \log_2\left(x_\mathrm{E} + \epsilon\mathrm{tr}\left(\bH_{\mathrm{E}}\bW\right)\right), f_4 = \log_2\left(x_\mathrm{E}\right)$ with $\bW = \bw\bw^H$, $x_\mathrm{J} = \left(1 - \epsilon\right)\mathrm{tr}\left(\bH_{\mathrm{J}}\bR\right) + \sigma_\mathrm{J}^2/P, \mathrm{J}\in\left\{\mathrm{B}, \mathrm{E}\right\}$.
%\begin{subequations}
%	\notag
%	\begin{align}
%		f_1 & = \log_2\left(\left(1 - \epsilon\right)\mathrm{tr}\left(\bH_{\mathrm{B}}\bR\right) + \epsilon\mathrm{tr}\left(\bH_{\mathrm{B}}\bw\bw^H \right) + \sigma_\mathrm{B}^2/P\right), \\[-4mm]
%		f_2 & = \log_2\left(\left(1 - \epsilon\right)\mathrm{tr}\left(\bH_{\mathrm{B}}\bR\right) + \sigma_\mathrm{B}^2/P\right), \\
%		f_3 & = \log_2\left(\left(1 - \epsilon\right)\mathrm{tr}\left(\bH_{\mathrm{E}}\bR\right) + \epsilon\mathrm{tr}\left(\bH_{\mathrm{E}}\bw\bw^H \right) + \sigma_\mathrm{E}^2/P\right), \\[-4mm]
%		f_4 & = \log_2\left(\left(1 - \epsilon\right)\mathrm{tr}\left(\bH_{\mathrm{E}}\bR\right) + \sigma_\mathrm{E}^2/P\right).
%	\end{align} 
%\end{subequations}

Due to the non-concave terms $- f_2$ and $-f_3$, the objective function in $\mathcal{P}_2$ is non-concave. We first obtain a concave lower bound of it by retaining the concave part and linearizing the non-concave part \cite{7776948,7776965}. To be specific, utilizing the following inequality
\begin{equation*} \label{ineq}
	\begin{aligned}
		g\left(\bX\right) = \ln\left(\mathrm{tr}\left(\bA\bX\right) + b\right) &\leq\mathrm{tr}\left(\left(\nabla g\left(\tilde{\bX}\right)\right)^H\left(\bX - \tilde{\bX}\right)\right) \\
		& + \ln\left(\mathrm{tr}\left(\bA\tilde{\bX}\right) + b\right),
	\end{aligned}
\end{equation*} 
where $\nabla g\left(\tilde{\bX}\right)$ represents the Euclidean gradient of $g$ at a given feasible point $\tilde{\bX}$, we can approximate $f_2$ and $f_3$ to linear functions $f_2^{'}$ and $f_3^{'}$ which are given by $f_2^{'} \!=\! \mathrm{tr}\left(\frac{\left(\log_2e\right)\left(1 - \epsilon\right)\bH_{\mathrm{B}}\left(\bR - \tilde{\bR}\right)}{\left(1 - \epsilon\right)\mathrm{tr}\left(\bH_{\mathrm{B}}\tilde{\bR}\right) + \sigma_\mathrm{B}^2/P}\right)  
\!+\!  \log_2\left(\left(1 - \epsilon\right)\mathrm{tr}\left(\bH_{\mathrm{B}}\tilde{\bR}\right) \right.$ $ + \sigma_\mathrm{B}^2/P\Big)$ and $f_3^{'} =  \mathrm{tr}\left(\frac{\left(\log_2e\right)\left(1 - \epsilon\right)\bH_{\mathrm{E}}\left(\bR - \tilde{\bR}\right)}{\left(1 - \epsilon\right)\mathrm{tr}\left(\bH_{\mathrm{E}}\tilde{\bR}\right) + \epsilon\mathrm{tr}\left(\bH_{\mathrm{E}}\bW \right) + \sigma_\mathrm{E}^2/P}\right)  
+  \log_2\left(\left(1 - \epsilon\right)\mathrm{tr}\left(\bH_{\mathrm{E}}\tilde{\bR}\right) \!+\! \epsilon\mathrm{tr}\left(\bH_{\mathrm{E}}\bW \right) \!+\! \sigma_\mathrm{E}^2/P\right)$.
%\begin{equation*}
%	\begin{aligned}
%		f_2^{'} = & \mathrm{tr}\left(\frac{\left(\log_2e\right)\left(1 - \epsilon\right)\bH_{\mathrm{B}}\left(\bR - \tilde{\bR}\right)}{\left(1 - \epsilon\right)\mathrm{tr}\left(\bH_{\mathrm{B}}\tilde{\bR}\right) + \sigma_\mathrm{B}^2/P}\right)   \\
%		+ & \log_2\left(\left(1 - \epsilon\right)\mathrm{tr}\left(\bH_{\mathrm{B}}\tilde{\bR}\right) + \sigma_\mathrm{B}^2/P\right),
%	\end{aligned}
%\end{equation*}
%\begin{equation*}
%	\begin{aligned}
%		f_3^{'} = & \mathrm{tr}\left(\frac{\left(\log_2e\right)\left(1 - \epsilon\right)\bH_{\mathrm{E}}\left(\bR - \tilde{\bR}\right)}{\left(1 - \epsilon\right)\mathrm{tr}\left(\bH_{\mathrm{E}}\tilde{\bR}\right) + \epsilon\mathrm{tr}\left(\bH_{\mathrm{E}}\bW \right) + \sigma_\mathrm{E}^2/P}\right)  \\
%		+ & \log_2\left(\left(1 - \epsilon\right)\mathrm{tr}\left(\bH_{\mathrm{E}}\tilde{\bR}\right) \!+\! \epsilon\mathrm{tr}\left(\bH_{\mathrm{E}}\bW \right) \!+\! \sigma_\mathrm{E}^2/P\right),
%	\end{aligned}
%\end{equation*}
Thus, the non-concave $C_s\left(\bR\right)$ can be lower bounded by a concave function that yields a surrogate objective function as follows:
\begin{equation}
	C_s\left(\bR\right) \geq C_s^{'}\left(\bR\right) = f_1 - f_2^{'} - f_3^{'} + f_4,
\end{equation}

Hence, the subproblem of optimizing $\bR$ can be rewritten as the following convex surrogate problem
\begin{equation*} \label{p_7}
	\mathcal{P}_3:
	\max_{\bR}~
	C_s^{'}\left(\bR\right)~~
	\mathrm{s.t.}~\eqref{p_c4}, \eqref{p_c3}, \eqref{cons_R}.
\end{equation*}
This is a semidefinite programming (SDP) problem which can be solved by existing software solvers, such as CVX \cite{8742603, 9364358}.

\subsubsection{Optimizing $\bQ$}
Then, we design the passive beamforming of RIS. With fixed $\epsilon$, $\bw$, and $\bR$, the problem of $\bQ$ can be expressed as 
\begin{equation*}	\label{p_4}
	\mathcal{P}_4:
	\max_{\bQ}~
	C_s\left(\bQ\right) = C_{\mathrm{B}} - C_{\mathrm{E}}~~
	~\mathrm{s.t.}~\eqref{p_c2}, \eqref{p_c3}.		
\end{equation*}

Define $\bU \!\triangleq\! \bu \bu^H$ with $\bu \!\!=\!\! t\left[\bq^T,1\right]^{T}, \left|t\right| \!\!=\!\! 1$, based on the identities $\mathrm{tr}\left(\bD^H\bA\bD\bB\right) \!\!=\!\! \bd^H\left(\bA\odot\bB^{T}\right)\bd$ and $\mathrm{tr}\left(\bD^H\bC\right) \!\!=\!\! \bd^H\bc$ with $\bD$ being a diagonal matrix, $\bd \!\!=\!\! \mathrm{Diag}\left(\bD\right)$ and $\bc \!\!=\!\! \mathrm{Diag}\left(\bC\right)$, terms $P_{\mathrm{cJ}}$ and $P_{\mathrm{rJ}}$ with $\mathrm{J}\!\in\!\left\{\mathrm{B},\mathrm{E}\right\}$ can be recast as
\begin{equation} \label{P_cJ}
	\begin{aligned}
	P_{\mathrm{cJ}} & = \epsilon P\left(\bq^H\bA_{\mathrm{cJ}}\bq + \bq^H\ba_{\mathrm{cJ}} + \ba_{\mathrm{cJ}}^H\bq + \bar{P}_{\mathrm{cJ}}\right) \\
	& =  \mathrm{tr}\left(\bR_{\mathrm{cJ}}\bU\right),
	\end{aligned}
\end{equation}
\begin{equation}	\label{P_rJ}
	\begin{aligned}
	P_{\mathrm{rJ}} & =  \left(1 - \epsilon\right)P\left(\bq^H\bA_{\mathrm{rJ}}\bq + \bq^H\ba_{\mathrm{rJ}} + \ba_{\mathrm{rJ}}^H\bq + \bar{P}_{\mathrm{rJ}}\right) \\
	& =  \mathrm{tr}\left(\bR_{\mathrm{rJ}}\bU\right),
	\end{aligned}
\end{equation}
where $\bA_{\mathrm{cJ}} = \left(\bh_{\mathrm{IJ}}\bh_{\mathrm{IJ}}^H\right)\odot\left(\left(\bH_{\mathrm{AI}}\bW\bH_{\mathrm{AI}}^H\right)^T\right)$, $\ba_{\mathrm{cJ}} = \mathrm{Diag}\left(\bh_{\mathrm{IJ}}\bh_{\mathrm{AJ}}^H\bW\bH_{\mathrm{AI}}^H\right)$, $\bar{P}_{\mathrm{cJ}} = \bh_{\mathrm{AJ}}^H\bW\bh_{\mathrm{AJ}}$. Similarly, $\bA_{\mathrm{rJ}} = \left(\bh_{\mathrm{IJ}}\bh_{\mathrm{IJ}}^H\right)\odot\left(\left(\bH_{\mathrm{AI}}\bR\bH_{\mathrm{AI}}^H\right)^T\right)$, $\ba_{\mathrm{rJ}} = \mathrm{Diag}\left(\bh_{\mathrm{IJ}}\bh_{\mathrm{AJ}}^H\bR\bH_{\mathrm{AI}}^H\right)$, $\bar{P}_{\mathrm{rJ}} = \bh_{\mathrm{AJ}}^H\bR\bh_{\mathrm{AJ}}$ and
\begin{equation}
	\notag
	\begin{aligned}
		\bR_{\mathrm{cJ}} = \epsilon P\left[\begin{array}{cc}
			\bA_{\mathrm{cJ}} & \ba_{\mathrm{cJ}} \\ \ba_{\mathrm{cJ}}^H & \bar{P}_{\mathrm{cJ}}
		\end{array}\right],
	 \bR_{\mathrm{rJ}} = \left(1 - \epsilon\right)P\left[\begin{array}{cc}
			\bA_{\mathrm{rJ}} & \ba_{\mathrm{rJ}} \\ \ba_{\mathrm{rJ}}^H & \bar{P}_{\mathrm{rJ}}
		\end{array}\right].
	\end{aligned}
\end{equation}

Based on the above derivations, the secrecy rate $C_s$ can be reformulated as $C_s\left(\tilde{\bu}\right) = f_5 - f_6 -f_7 + f_8$, where 
\begin{equation}
	\notag
	\begin{aligned}
		f_5 & = \log_2\left(\mathrm{tr}\left(\left(\bR_{\mathrm{cB}} + \bR_{\mathrm{rB}}\right)\bU\right) + \sigma_\mathrm{B}^2\right) = \log_2\left(\tilde{\bu}^H\br_1\right), \\
		f_6 & = \log_2\left(\mathrm{tr}\left(\bR_{\mathrm{rB}}\bU\right) + \sigma_\mathrm{B}^2\right) = \log_2\left(\tilde{\bu}^H\br_2\right), \\
		f_7 & = \log_2\left(\mathrm{tr}\left(\left(\bR_{\mathrm{cE}} + \bR_{\mathrm{rE}}\right)\bU\right) + \sigma_\mathrm{E}^2\right) = \log_2\left(\tilde{\bu}^H\br_3\right), \\
		f_8 & = \log_2\left(\mathrm{tr}\left(\bR_{\mathrm{rE}}\bU\right) + \sigma_\mathrm{E}^2\right)= \log_2\left(\tilde{\bu}^H\br_4\right),
	\end{aligned}
\end{equation}
where $\tilde{\bu} \!\!\!=\!\!\! \left[1,\bar{\bu}^T\right]^T$ with $\bar{\bu} \!\!\!=\!\!\! \mathrm{vec}\left(\bU\right)$, $\br_1 \!\!=\!\! \left[\sigma_\mathrm{B}^2,\mathrm{vec}^T\left(\bR_{\mathrm{cB}} + \bR_{\mathrm{rB}}\right)\right]^T$, $\br_2 = \left[\sigma_\mathrm{B}^2,\mathrm{vec}^T\left(\bR_{\mathrm{rB}}\right)\right]^T$, $\br_3 = \left[\sigma_\mathrm{E}^2,\mathrm{vec}^T\left(\bR_{\mathrm{cE}} + \bR_{\mathrm{rE}}\right)\right]^T$, and $\br_4 = \left[\sigma_\mathrm{E}^2,\mathrm{vec}^T\left(\bR_{\mathrm{rE}}\right)\right]^T$.

{{Note that $C_s\left(\tilde{\bu}\right)$ is non-concave due to the non-concave part $-f_6$ and $-f_7$. Similarly, by applying the first-order Taylor approximation to $f_6$ and $f_7$, we can obtain a concave lower bound of $C_s\left(\tilde{\bu}\right)$ which is given as follows
\begin{equation}
	C_s\left(\tilde{\bu}\right) \geq C_s^{'}\left(\tilde{\bu}\right) = f_5 - f_6^{'} - f_7^{'} + f_8,
\end{equation}}}
where $f_6^{'} = \left(\log_2e\right)\br_2^H\left(\tilde{\bu} - \tilde{\bu}_0\right)/\left(\tilde{\bu}_0^H\br_2\right) + \log_2\left(\tilde{\bu}_0^H\br_2\right)$ and $f_7^{'} = \left(\log_2e\right)\br_3^H\left(\tilde{\bu} - \tilde{\bu}_0\right)/\left(\tilde{\bu}_0^H\br_3\right) + \log_2\left(\tilde{\bu}_0^H\br_3\right)$ with $\tilde{\bu}_0$ being a feasible value of $\tilde{\bu}$.
%\begin{equation*}
%	\begin{aligned}
%	f_6^{'} = \left(\log_2e\right)\br_2^H\left(\tilde{\bu} - \tilde{\bu}_0\right)/\left(\tilde{\bu}_0^H\br_2\right) + \log_2\left(\tilde{\bu}_0^H\br_2\right), %\frac{\left(\log_2e\right)\br_2^H\left(\tilde{\bu} - \tilde{\bu}_0\right)}{\tilde{\bu}_0^H\br_2}  + \log_2\left(\tilde{\bu}_0^H\br_2\right) 
%	\end{aligned}
%\end{equation*}
%\begin{equation*}
%	\begin{aligned}
%		f_7^{'} = \left(\log_2e\right)\br_3^H\left(\tilde{\bu} - \tilde{\bu}_0\right)/\left(\tilde{\bu}_0^H\br_3\right) + \log_2\left(\tilde{\bu}_0^H\br_3\right), %\frac{\left(\log_2e\right)\br_3^H\left(\tilde{\bu} - \tilde{\bu}_0\right)}{\tilde{\bu}_0^H\br_3}  + \log_2\left(\tilde{\bu}_0^H\br_3\right)
%	\end{aligned}
%\end{equation*}

{{Next, we transform the radar detection constraint into a more tractable form.}} Define $\bE = \left[\left(\mathrm{diag}\left(\bh_{\mathrm{IE}}^H\right)\bH_{\mathrm{AI}}\right)^T,\bh_{\mathrm{AE}}^*\right]$, the numerator in $\gamma_{\mathrm{A}}$ can be rewritten as
\begin{equation} \label{radar_power_A}
	\begin{aligned}
		\left(1 - \epsilon\right)P\mathrm{tr}\left(\bH_{\mathrm{E}}^H\bH_{\mathrm{E}}\bR\right) = & \mathrm{tr}\left(\|\bh_{\mathrm{E}}\|^2\bH_{\mathrm{E}}\bR\right) \\
		= & \mathrm{tr}\left(\bu^H\bE^H\bE\bu\bu^H\bR_{\mathrm{rE}}\bu\right) \\
		= & \mathrm{tr}\left(\bE^H\bE\bU\bR_{\mathrm{rE}}\bU\right) \\
		= & \bar{\bu}^H\bG_{\mathrm{rE}}\bar{\bu},
	\end{aligned}
\end{equation}
where $\bG_{\mathrm{rE}} = \bR_{\mathrm{rE}}^T\otimes \left(\bE^H\bE\right)$ \cite{9364358}. Similarly, the denominator in $\gamma_{\mathrm{A}}$ can be expressed as 
\begin{equation}
	\begin{aligned}
		\epsilon P \bw^H\bH_{\mathrm{E}}^H\bH_{\mathrm{E}}\bw = \bar{\bu}^H\bG_{\mathrm{cE}}\bar{\bu},
	\end{aligned}
\end{equation} 
where $\bG_{\mathrm{cE}} = \bR_{\mathrm{cE}}^T\otimes \left(\bE^H\bE\right)$. Thus, $\gamma_{\mathrm{A}}$ can be rewritten as
\begin{equation}
	\gamma_{\mathrm{A}} = \frac{\bar{\bu}^H\bG_{\mathrm{rE}}\bar{\bu}}{\bar{\bu}^H\bG_{\mathrm{cE}}\bar{\bu} + N\sigma_{\mathrm{A}}^2},
\end{equation}
and the radar constraint \eqref{p_c3} can be further expressed as
\begin{equation} \label{p_c3_trans}
	\tilde{\bu}^H \bK \tilde{\bu} \leq 0,
\end{equation}
where $\bK = \left[\begin{array}{cc}
	\gamma N \sigma_{\mathrm{A}}^2 & \bf{0} \\ \bf{0} & \bK_1
\end{array}\right]$ with $\bK_1 = \gamma\bG_{\mathrm{cE}} - \bG_{\mathrm{rE}}$. {{Since $\bK$ is indefinite, constraint \eqref{p_c3_trans} is non-convex. Note that $\bK$ is an Hermitian matrix which can be expressed as the sum of one positive semidefinite matrix and one negative semidefinite matrix, i.e., $\bK = \bK^{+} + \bK^{-}$.}} We can rewrite \eqref{p_c3_trans} as
\begin{equation} \label{po_ne_K}
	\tilde{\bu}^H \bK^{+} \tilde{\bu} + \left(\tilde{\bu}^H \bK^{-} \tilde{\bu}\right) \leq 0.
\end{equation}  
 
For any vector $\bz\in\mathbb{C}^{\left(M + 1\right)^2 + 1}$, $\left(\tilde{\bu} - \bz\right)^H\bK^{-}\left(\tilde{\bu} - \bz\right) \leq 0$. Let $\bz = \tilde{\bu}_0$ and expand the left side, we can obtain $\tilde{\bu}^H \bK^{-} \tilde{\bu} \leq 2\mathfrak{R}\left(\tilde{\bu}_0^H\bK^{-}\tilde{\bu}\right) - \tilde{\bu}_0^H\bK^{-}\tilde{\bu}_0$. Thus, we approximate \eqref{p_c3_trans} by  
\begin{equation} \label{p_c3_trans2}
	\tilde{\bu}^H \bK^{+} \tilde{\bu} + 2\mathfrak{R}\left(\tilde{\bu}_0^H\bK^{-}\tilde{\bu}\right) - \tilde{\bu}_0^H\bK^{-}\tilde{\bu}_0 \leq 0.
\end{equation}

Via the above manipulations, a suboptimal solution to problem $\mathcal{P}_4$ can be obtained by solving
\begin{subequations}
	\begin{align}	\label{p_5}
		\mathcal{P}_5: 
		\max_{\tilde{\bu}}~  
		& C_s^{'}\left(\tilde{\bu}\right)\\
		~\mathrm{s.t.}
		\notag
		& ~  \eqref{p_c3_trans2}, \\
		& ~ \tilde{\bu}_1 = 1, \bU_{i,i} = 1, \forall i = 1,\dots,M + 1, \\
		& ~ \bU = \bu\bu^H, \mathrm{rank}\left(\bU\right) = 1,
	\end{align}
\end{subequations}
where $\tilde{\bu}_1$ represents the first element of $\tilde{\bu}$.
This is a semidefinite relaxation problem. Omitting the rank-one constraint, it becomes a standard SDP problem which can be solved by CVX. Then, the Gaussian randomization can be used to recover the rank-one solution \cite{9199556,9438645}. After obtaining $\bu$, the optimal $\bq$ can be calculated by $\bq^{*} = \frac{\bu_{\left[1:M\right]}}{\bu_{M + 1}}$, $\bQ^{*} = \mathrm{diag}\left(\bq^{*}\right)$. 

\subsubsection{Optimizing $\epsilon$}
With given $\bw$, $\bR$, and $\bQ$, the subproblem related to the power allocation factor $\epsilon$ is expressed as follows.
\begin{equation*}
	\begin{aligned}	
	\mathcal{P}_6:
			\max_{\epsilon}~
			C_s\left(\epsilon\right) = C_{\mathrm{B}} - C_{\mathrm{E}}
			~~~\mathrm{s.t.}
			~~\eqref{p_c_ep}, \eqref{p_c3}.
		\end{aligned}
\end{equation*}

Optimal $\epsilon$ to the problem can be efficiently obtained by one-dimensional search within $\left[0,1\right]$. We summarize the whole ZF-based BCD algorithm in Algorithm \ref{BCD}. 

\emph{Complexity Analysis:} The overall computational complexity of Algorithm \ref{BCD} is determined by solving SDP problems of $\mathcal{P}_3$ and $\mathcal{P}_5$. According to \cite{2002Interior}, the computational complexity of solving a SDP problem with $m$ constraints where each constraint involves a $n\times n$ positive semidefinite matrix is $\mathcal{O}\left(mn^3 + m^2n^2 + m^3\right)$ in each iteration. For problems $\mathcal{P}_3$ and $\mathcal{P}_5$, we have $m \!=\! 3, n \!=\! N$ and $m \!=\! M + 2, n \!=\! M + 1$, respectively. Hence, the overall computational complexity of Algorithm \ref{BCD} is about $\mathcal{O}\left(I_{\mathrm{iter}}\Big(\left.3N^3 \!+\! 9N^2 \!+\! 27 \!+\! \left(M + 2\right)\right.\right.$ 
$\left.\left. \left(M \!+\! 1\right)^3 \!+\! \left(M \!+\! 2\right)^2\left(M \!+\! 1\right)^2 \!+\! \left(M \!+\! 2\right)^3\right)\right)$ where $I_{\mathrm{iter}}$ denotes the total number of iterations.       
% $\left.\left.\sqrt{M + 1}\log\left(2\left(M + 1\right)^4 \!+\! \left(M + 1\right)^3\right)\right)\right)$
 \begin{algorithm}[htbp]
	\caption{\it ZF-based BCD Algorithm for Solving $\mathcal{P}_1$}
	\label{BCD}
	\begin{algorithmic} 
		\State 1. Set $\epsilon_{\mathrm{min}} = 0$, $\epsilon_{\mathrm{max}} = 1$, $\epsilon_s > 0$ with $\epsilon_s$ being the search step, and convergence accuracy $\epsilon_b$.
		\State 2. Set initial points $\epsilon_0 \!=\! \epsilon_s$, $\bw_0$, $\bR_0$, $\bQ_0$, and iteration number $i = 0$. Calculate $C_{s_0} = C_s\left(\epsilon_0,\bw_0,\bR_0,\bQ_0\right)$.
		\Repeat
		\State 3. Set $ i = i + 1$
		\State 4. Calculate $\bw_i$ for given $\epsilon_{i - 1}$, $\bR_{i - 1}$, and $\bQ_{i - 1}$  by \eqref{zf_w}.
		\State 5. With given $\bw_i$, $\epsilon_{i - 1}$, and $\bQ_{i - 1}$, obtain $\bR_i$ by solving problem $\mathcal{P}_3$. 
		\State 6. Obtain $\bQ_i$ with given $\epsilon_{i - 1}$, $\bw_i$ and $\bR_i$ by solving problem $\mathcal{P}_5$. 
		\State 7. Obtain $\epsilon_i$ with fixed $\bw_i$, $\bR_i$, and $\bQ_i$ by solving problem $\mathcal{P}_6$. 
		\State 8. Calculate $C_{s_i} = C_s\left(\epsilon_i, \bw_i,\bR_i,\bQ_i\right)$.   
		\Until 
		$\frac{\left|C_{s_i} -C_{s_{i - 1}}\right|}{\left|C_{s_{i - 1}}\right|}\leq \epsilon_b$.
		\State 9. Output $\left(\epsilon_i, \bw_i,\bR_i,\bQ_i\right)$ as the solution to problem $\mathcal{P}_1$.
%		\State 9. If $R_{s_i} > R_s$, $R_s = R_{s_i}$. Update $\epsilon = \epsilon + \epsilon_s$. 
%		\Until 
%		$\epsilon + \epsilon_s > \epsilon_{max}$.
%		\State 10. Output $R_s$ and the corresponding $\left(\epsilon,\bw,\bR,\bQ\right)$ as the solution to problem $\mathcal{P}$. 
	\end{algorithmic}
\end{algorithm} 
 
\section{Secure communication in DFRC System}
In this section, we investigate the DFRC system as also shown in Fig. \ref{fig1}. Here, Alice transmits shared signals for both radar detection and communication. Two algorithms, i.e., the Dinkelbach method-based AO algorithm and the RCG-based AO algorithm, are developed to maximize the secrecy rate subjected to the radar detection constraint.  

\subsection{Problem Formulation}
In DFRC system, the received signals at Bob and Eve can be expressed as
\begin{equation} \label{y_B}
	y_{\mathrm{B}} = \sqrt{P}\bh_{\mathrm{B}}\bw x + n_{\mathrm{B}},
\end{equation} 
\begin{equation} \label{y_E}
	y_{\mathrm{E}} = \sqrt{P}\bh_{\mathrm{E}}\bw x + n_{\mathrm{E}},
\end{equation}
% \left(\bh_{\mathrm{IB}}\bQ\bH_{\mathrm{AI}} + \bh_{\mathrm{AB}} \right)
% \left(\bh_{\mathrm{IE}}\bQ\bH_{\mathrm{AI}} + \bh_{\mathrm{AE}} \right)
respectively, where $x \!\in\! \mathbb{C}, \mathbb{E}\left[xx^H\right] = 1$ is a scalar denoting the transmitted signal used for both radar detection and communication;
% $\mathbb{E}\left[xx^H\right] = 1$ and 
$\bw \in \mathbb{C}^{N\times 1}$ is the transmit beamforming vector. Recall that $\bh_{\mathrm{B}} = \bh_{\mathrm{IB}}^H\bQ\bH_{\mathrm{AI}} + \bh_{\mathrm{AB}}^H$ and $\bh_{\mathrm{E}} = \bh_{\mathrm{IE}}^H\bQ\bH_{\mathrm{AI}} + \bh_{\mathrm{AE}}^H$ represent the effective channel of Alice-Bob and Alice-Eve, respectively. Thus, the achievable rate of Bob and Eve can be expressed as
\begin{equation} \label{R_B}
	C_\mathrm{B} = \log_2\left(1 +\frac{P_\mathrm{B}}{\sigma_\mathrm{J}^2}\right) = \log_2\left(1 + \frac{P\bh_{\mathrm{B}}\bw \bw^H\bh_{\mathrm{B}}^H}{\sigma_\mathrm{B}^2}\right),
\end{equation}
\begin{equation} \label{R_E}
	C_\mathrm{E} =  \log_2\left(1 +\frac{P_\mathrm{E}}{\sigma_\mathrm{E}^2}\right) = \log_2\left(1 + \frac{P\bh_{\mathrm{E}}\bw \bw^H\bh_{\mathrm{E}}^H}{\sigma_\mathrm{E}^2}\right),
\end{equation}
respectively, where $P_\mathrm{B}$ and $P_\mathrm{E}$ denote the received useful signal power at Bob and Eve. 

The received echo signals at Alice can be expressed as
\begin{equation}
	\by_\mathrm{A} = \sqrt{P}\bH_{\mathrm{E}}\bw x + \bn_{\mathrm{A}},
\end{equation}
and thus the SNR at Alice can be calculated as
\begin{equation}
	\gamma_{\mathrm{A}} = \frac{P\bw^H\bH_{\mathrm{E}}^H\bH_{\mathrm{E}}\bw}{N\sigma_{\mathrm{A}}^2}.
\end{equation}

The secrecy rate is supposed to be maximized while satisfying the radar sensing constraint which is formulated as
\begin{subequations}
	\begin{align}	\label{p}
		\mathcal{P}_7:
		\max_{\bw,\bQ}~
		&C_s\left(\bw,\bQ\right) = C_{\mathrm{B}} - C_{\mathrm{E}}\\
		~\mathrm{s.t.}
		\label{pc1}
		~~& \|\bw\| = 1,\\
		\label{pc2}
		~~& \left|\bQ_{i,i} \right| = 1, \forall i=1,2,\dots,M,\\
		\label{pc3}
		~~& \gamma_{\mathrm{A}}  \geq \gamma.
	\end{align}
\end{subequations}
It is an intractable non-convex problem due to the non-convex unit modulus constraint \eqref{pc2} and the non-convex constraint \eqref{pc3} with two optimization variables highly coupled. 

\subsection{Problem Solution}
{{In this subsection, by exploiting the structures of the objective function and optimization variables, we propose two algorithms, i.e., the Dinkelbach method-based AO algorithm and the RCG-based AO algorithm, to solve the problem $\mathcal{P}_7$.}}
\subsubsection{Dinkelbach Method-based AO Algorithm}
We first decompose problem $\mathcal{P}_7$ into two subproblems related to $\bw$ and $\bQ$, respectively. Then, the Dinkelbach method is exploited to address each subproblem. The key idea of AO algorithm is to alternately optimize $\bw$ and $\bQ$ until the objective function converges such that a suboptimal solution to problem $\mathcal{P}_7$ can be obtained. Next, we introduce the steps in details.  
  
It is easily observed that problem $\mathcal{P}_7$ can be equivalently expressed as
\begin{equation*}	\label{p'}
		\mathcal{P}_7^{'}:
		\max_{\bw,\bQ}~  
		\frac{1 + P_{\mathrm{B}}/\sigma_\mathrm{B}^2}{1 + P_{\mathrm{E}}/\sigma_\mathrm{E}^2}~~
		~\mathrm{s.t.}
		~ \eqref{pc1}-\eqref{pc3}.
\end{equation*}

%This is a fractional programming problem which can be solved by Dinkelbach method \cite{9199556}. Specifically, by introducing an auxiliary variable $\mu$, problem $\mathcal{P}_7^{'}$ can be equivalently transformed to
%\begin{equation} \label{p1}
%	\begin{aligned}	
%		\mathcal{P}_8: 
%		\max_{\bw,\bQ}~  
%		&\left(1 + P_B/\sigma_\mathrm{B}^2\right) - \mu\left(1 + P_E/\sigma_\mathrm{E}^2\right)\\
%		~\mathrm{s.t.}
%		& ~ \eqref{pc1}-\eqref{pc3},
%	\end{aligned}
%\end{equation}
%where $\mu$ in the $i$-th iteration is calculated by $\mu_i = \left[\frac{1 + P_B/\sigma_\mathrm{B}^2}{1 + P_E/\sigma_\mathrm{E}^2}\right]_{i - 1}$. Then, the AO algorithm, i.e., optimizing $\bw$ and $\bQ$ alternately until the OF converges, is used to solve $\mathcal{P}_8$.

%\subsubsection{optimization of $\bw$}

With fixed $\bQ$, the subproblem of $\bw$ is formulated as
\begin{equation*}	\label{p2}
		\mathcal{P}_8: 
		\max_{\bw}~  
		\frac{1 + P_{\mathrm{B}}/\sigma_\mathrm{B}^2}{1 + P_{\mathrm{E}}/\sigma_\mathrm{E}^2}~~
		~\mathrm{s.t.}
		 ~ \eqref{pc1}, \eqref{pc3}.
\end{equation*}

Define $\bW \triangleq \bw \bw^H$, problem $\mathcal{P}_8$ can be further cast as
\begin{subequations}
	\begin{align}	\label{p3}
		\mathcal{P}_{9}: 
		\max_{\bW}~  
		& \frac{1 + \mathrm{tr}\left( \bM_{\mathrm{B}}\bW\right)}{1 + \mathrm{tr}\left( \bM_{\mathrm{E}}\bW\right)} \\
		~\mathrm{s.t.}
		\label{p3_c1}
		& ~ \mathrm{tr}\left(\bW\right) = 1, \\
		\label{p3_c2}
		& ~ \mathrm{tr}\left(\bM_{\mathrm{E}}^{'} \bW\right) \geq \gamma N\sigma_{\mathrm{A}}^2, \\
		\label{p3_c3}
		& ~ \bW = \bW^H, \bW \succeq {\bf 0 }, \mathrm{rank}\left(\bW\right) = 1,
	\end{align}
\end{subequations}
where $\bM_{\mathrm{J}} = P/\sigma_{\mathrm{J}}^2\bH_{\mathrm{J}}$ with $\mathrm{J} \in \{\mathrm{B}, \mathrm{E}\}$ and $\bM_{\mathrm{E}}^{'} = P\bH_{\mathrm{E}}^H\bH_{\mathrm{E}}$.

{{To facilitate the Dinkelbach method, we first omit the rank-one constraint. By introducing an auxiliary variable $\mu$, problem $\mathcal{P}_9$ without the rank-one constraint can be equivalently transformed into \cite{9199556}
\begin{equation} \label{p1}
	\notag
	\begin{aligned}	
			\mathcal{P}_{10}: 
			\max_{\bW}~  
			&\mathrm{tr}\left(\left(\bM_{\mathrm{B}} - \mu\bM_{\mathrm{E}}\right)\bW\right)\\
			~\mathrm{s.t.}
			& ~ \bW = \bW^H, \bW \succeq {\bf 0}, \eqref{p3_c1}, \eqref{p3_c2},
		\end{aligned}
\end{equation}
where $\mu$ in the $i$-th iteration is updated by $\mu_i = \frac{1 + P_{\mathrm{B}}\left(\bW_{i - 1}\right)/\sigma_\mathrm{B}^2}{1 + P_{\mathrm{E}}\left(\bW_{i - 1}\right)/\sigma_\mathrm{E}^2}$. In each iteration, we can employ CVX to solve the SDP problem $\mathcal{P}_{10}$. When $\mu$ converges, $\bW$ is treated as the solution.}} Then, Gaussian randomization can be exploited to recover a rank-one solution.    

%\begin{equation}	\label{M_B}
%	\notag
%	\begin{aligned}
%	\bM_J   = &P/\sigma_J^2\left(\bH_{\mathrm{AI}}^H\bQ^H\bh_{\mathrm{IJ}}^H\bh_{\mathrm{IJ}}\bQ\bH_{\mathrm{AI}} + 
%	\bH_{\mathrm{AI}}^H\bQ^H\bh_{\mathrm{IJ}}^H\bh_{\mathrm{AJ}}  \right. \\
%	& \left.  +~\bh_{\mathrm{AJ}}^H\bh_{\mathrm{IJ}}\bQ\bH_{\mathrm{AI}} + \bh_{\mathrm{AJ}}^H\bh_{\mathrm{AJ}}\right), J\in\left\{B,E\right\},
%	\end{aligned}
% \end{equation}

% \subsubsection{optimization of $\bQ$}
On the other hand, with fixed $\bw$, the subproblem of optimizing $\bQ$ can be formulated as
\begin{equation} 	\label{p4}
		\mathcal{P}_{11}: 
		\max_{\bQ}~  
		\frac{1 + P_{\mathrm{B}}/\sigma_\mathrm{B}^2}{1 + P_{\mathrm{E}}/\sigma_\mathrm{E}^2}~~
		~\mathrm{s.t.}
		 ~ \eqref{pc2}, \eqref{pc3}.
\end{equation}

Similar to \eqref{P_cJ} and \eqref{P_rJ}, $P_{\mathrm{J}}$ with $\mathrm{J} \in \{\mathrm{B}, \mathrm{E}\}$ in \eqref{p4} can be rewritten as
\begin{equation}
	\begin{aligned}
	P_{\mathrm{J}} &= P\left(\bq^H\bF_{\mathrm{J}}\bq + \bq^H\bm{f}_{\mathrm{J}} + \bm{f}_{\mathrm{J}}^H\bq + g_{\mathrm{J}}\right) \\
	& = \mathrm{tr}\left(\bR_{\mathrm{J}}\bU\right) = \bar{\bu}^H\br_{\mathrm{J}}, \\
	\end{aligned}
\end{equation} 
%\begin{equation}
%	\begin{aligned}
%	P_{\mathrm{E}} & = P\left(\bq^H\bF_{\mathrm{E}}\bq + \bq^H\bm{f}_{\mathrm{E}} + \bm{f}_{\mathrm{E}}^H\bq + g_{\mathrm{E}}\right)\\
%	& = \mathrm{tr}\left(\bR_{\mathrm{E}}\bU\right) = \bar{\bu}^H\br_{\mathrm{E}},
%	\end{aligned}
%\end{equation}
where $g_{\mathrm{J}} = \mathrm{tr}\left(\bh_{\mathrm{AJ}}\bh_{\mathrm{AJ}}^H\bW\right)$, $\bm{f}_{\mathrm{J}} = \mathrm{Diag}\left(\bh_{\mathrm{IJ}}\bh_{\mathrm{AJ}}^H\bW\bH_{\mathrm{AI}}^H\right)$, $\bF_{\mathrm{J}} \!\!=\!\! \left(\bh_{\mathrm{IJ}}\bh_{\mathrm{IJ}}^H\right)\odot\left(\left(\bH_{\mathrm{AI}}\bW\bH_{\mathrm{AI}}^H\right)^T\right)$, $\bar{\bu} \!\!=\!\! \mathrm{vec}\left(\bU\right)$, $\bR_{\mathrm{J}} = P
\left[\begin{array}{cc}
	\bF_{\mathrm{J}} & \bm{f}_{\mathrm{J}} \\ \bm{f}_{\mathrm{J}}^H & g_{\mathrm{J}}
\end{array}\right]$, $\br_{\mathrm{J}} \!=\! \mathrm{vec}\left(\bR_{\mathrm{J}}\right)$. 

Also, the radar constraint \eqref{pc3} can be reformulated as 
\begin{equation} \label{S_E}
\gamma_{\mathrm{A}} = \bar{\bu}^H\bS_{\mathrm{E}}\bar{\bu} \geq \gamma,
\end{equation}
where $\bS_{\mathrm{E}} = \frac{1}{N\sigma_{\mathrm{A}}^2}\bR_{\mathrm{E}}^T\otimes\left(\bE^H\bE\right)$. One can prove that $\bS_{\mathrm{E}}$ is a positive semidefinite matrix. Reviewing the derivation of \eqref{p_c3_trans2}, a convex subset of constraint \eqref{S_E} can be established as
\begin{equation}
	\bar{\bu}^H\bS_{\mathrm{E}}\bar{\bu} \geq 2\mathfrak{R}\left(\bar{\bu}_0^H\bS_{\mathrm{E}}\bar{\bu}\right) - \bar{\bu}_0^H\bS_{\mathrm{E}}\bar{\bu}_0 \geq \gamma,
\end{equation}
where $\bar{\bu}_0$ being a feasible value of $\bar{\bu}$. 
Then, a suboptimal solution to problem $\mathcal{P}_{11}$ can be acquired by tackling the following problem according to the steps of solving $\mathcal{P}_9$.
\begin{subequations}
	\begin{align}	\label{sec_4_p5}
		\mathcal{P}_{12}: 
		\max_{\bar{\bu}}~  
		& \frac{1 + \bar{\bu}^H\br_{\mathrm{B}}/\sigma_\mathrm{B}^2}{1 + \bar{\bu}^H\br_{\mathrm{E}}/\sigma_\mathrm{E}^2} \\
		~\mathrm{s.t.}
		& ~ \bU_{i,i} = 1, \forall i = 1,\dots,M + 1, \\
		& ~ \bU = \bu\bu^H, \mathrm{rank}\left(\bU\right) = 1,\\
		& ~ 2\mathfrak{R}\left(\bar{\bu}_0^H\bS_{\mathrm{E}}\bar{\bu}\right) - \bar{\bu}_0^H\bS_{\mathrm{E}}\bar{\bu}_0 \geq \gamma. 
	\end{align}
\end{subequations}

\subsubsection{RCG-based AO Algorithm}
{{In the Dinkelbach method-based AO algorithm, the problem of optimizing $\bw$ or $\bQ$ is ultimately transformed into the SDP problem which incurs high computational complexity \cite{8288677}. Inspired by the fact that the feasible sets of $\bw$ and $\bq$ are a complex hypersphere and an oblique manifold, respectively, a low-complexity algorithm, namely, the RCG-based AO algorithm, is proposed to handle problem $\mathcal{P}_7^{'}$. It searches $\bw$ or $\bq$ directly over their feasible sets, which is more computationally efficient.}}

Denote the feasible sets of $\bw$ and $\bq$ as 
\begin{subequations}
	\begin{align}
		\mathcal{S} & = \left\{\left.\bw \in \mathbb{C}^{N \times 1}\right|\left\|\bw\right\| = 1 \right\}, \\ 
		\mathcal{O} & = \left\{\bq\in\mathbb{C}^{M \times 1}\left|\left[\bq\bq^H\right]_{i,i}=1,\forall i = 1,2,\dots,M\right.\right\},
	\end{align}
\end{subequations}
respectively \cite{article1}.
To apply the RCG algorithm, we first incorporate the constraint \eqref{pc3} as a penalty term into the objective function in problem $\mathcal{P}_7^{'}$ as follow \cite{8288677}:
\begin{equation} \label{penalty_p} 
	\min_{\bw,\bq}~  
	{\frac{1 + P_{\mathrm{E}}/\sigma_\mathrm{E}^2}{1 + P_{\mathrm{B}}/\sigma_\mathrm{B}^2} + \zeta\mathrm{max}\{0,\gamma - \gamma_{\mathrm{A}}\}},
\end{equation} 
where $\zeta \gg 1$ is a penalty factor. When applying the RCG algorithm, the objective function should be differentiable and thus we can obtain its gradient. However, the $\mathrm{max}$ function in \eqref{penalty_p} is non-smooth and its gradient does not exist. Hence, we approximate it by a smooth function as follows
\begin{equation} \label{max_approximate}
	\mathrm{max}\{0,\gamma - \gamma_{\mathrm{A}}\} \leq \epsilon_1 \ln\left(1 + e^{\frac{\gamma - \gamma_{\mathrm{A}}}{\epsilon_1}}\right),
\end{equation}
where $\epsilon_1$ is a small positive number \cite{8288677}. Thus, the secrecy rate maximization problem is transformed into 
\begin{subequations}
	\begin{align}	\label{p6}
		\mathcal{P}_{13}: 
		\min_{\bw,\bq}~ \! 
		&f = {\frac{1 + P_{\mathrm{E}}/\sigma_\mathrm{E}^2}{1 + P_{\mathrm{B}}/\sigma_\mathrm{B}^2} + \zeta \epsilon_1 \ln\left(1 + e^{\frac{\gamma - \gamma_{\mathrm{A}}}{\epsilon_1}}\right)}\\
		~\mathrm{s.t.}~& \bw \in \mathcal{S}, \bq \in \mathcal{O}. 
	\end{align}
\end{subequations} 

Next, the RCG algorithm is adopted to optimize $\bw$ and $\bq$ alternately.  
Given $\bq$, the problem of $\bw$ is expressed as
\begin{subequations}
	\begin{align}	\label{p7}
		\!\!\!\!\!\!\!\mathcal{P}_{14}\!:\!
		\min_{\bw} 
		&{f\!\left(\bw\right) \!=\! \frac{1 \!+\! \bw^H\bM_{\mathrm{E}}\bw}{1 \!+\! \bw^H\bM_{\mathrm{B}}\bw} \!+\! \zeta\epsilon_1 \!\ln\!\!\left(\!\!1 \!+\! e^{\frac{\gamma  - \frac{\bw^H\bM_{\mathrm{E}}^{'}\bw}{N\sigma_{\mathrm{A}}^2}}{\epsilon_1}}\!\right)}\\
		\mathrm{s.t.}~& \bw \in \mathcal{S}.	
	\end{align}
\end{subequations}
According to the RCG algorithm, given the $k$-th iteration solution, $\bw_k$, the next point $\bw_{k + 1}$ can be calculated as 
\begin{equation}	
	\begin{aligned}		\label{w_1}
		\bw_{k+1}  = R_{\bw_k}\left(s_k\bm{\eta}_k\right) 
		 = \frac{\bw_k + s_k\bm{\eta}_k}{\|\bw_k + s_k\bm{\eta}_k\|},
	\end{aligned}
\end{equation} 
where $R$ represents a retraction on $\mathcal{S}$, $s_k \geq 0$ is the step size, and $\bm{\eta}_k$ is the search direction which can be calculated by 
\begin{equation} \label{eta_k}
	\bm{\eta}_k = -\mathrm{grad} f\left(\bw_k\right) + \mu_k \mathcal{T}_{\bw_{k-1}}\left(\bm{\eta}_{k-1}\right),
\end{equation} 
where $\mu_k \geq 0$.
%and different choices of $\mu_k$ have been proposed, such as Fletcher and Reeves parameter and Polak-Ribi$\grave{\mathrm{e}}$re parameter, and so on \cite{Numericaloptimization}; 
Operator $\mathrm{grad} f\left(\bw_k\right)$ denotes the Riemannian gradient of $f$ at point $\bw_k$, which is calculated based on the orthogonal projection of the Euclidean gradient $\nabla f\left(\bw_k\right)$ onto the tangent space $T_{\bw_k}\mathcal{S}$. The Euclidean gradient of $f\left(\bw\right)$ at point $\bw_k$ is given as follows
\begin{equation}
	\begin{aligned} \label{nablaf}
		\nabla f\left(\bw_k\right) = &
		2\left\{\frac{\bM_{\mathrm{E}}\bw_k}{1 \!+\! \bw_k^H\bM_{\mathrm{B}}\bw_k} - \frac{\bM_{\mathrm{B}}\bw_k\left(1 \!+\! \bw_k^H\bM_{\mathrm{E}}\bw_k\right)}{\left(1 \!+\! \bw_k^H\bM_{\mathrm{B}}\bw_k\right)^2}  \right.\\
		- & \left. \frac{\zeta  \bM_{\mathrm{E}}^{'}\bw_k}{N\sigma_{\mathrm{A}}^2 \left(1\!+\! e^{\frac{\bw_k^H\bM_{\mathrm{E}}^{'}\bw_k - \gamma N\sigma_{\mathrm{A}}^2}{\epsilon_1 N\sigma_{\mathrm{A}}^2}}\right)}\right\}.
	\end{aligned}
\end{equation}

For a complex hypersphere, the tangent space at point $\bw_k$ is expressed as $T_{\bw_k}\mathcal{S} = \left\{\bv \in \mathbb{C}^{N \times 1}\mid \mathfrak{R}\left(\bw_k^H\bv\right) = 0 \right\}$ \cite{article1}.
%\begin{equation} \label{TwS}
%	T_{\bw_k}\mathcal{S} = \left\{\bv \in \mathbb{C}^{N \times 1}\mid \mathfrak{R}\left(\bw_k^H\bv\right) = 0 \right\}.
%\end{equation}
Thus, the Riemannian gradient $\mathrm{grad} f\left(\bw_k\right)$ is calculated as
\begin{equation} \label{gradf}
	\mathrm{grad}f\left(\bw_k\right) = \nabla f\left(\bw_k\right) - \mathfrak{R}\left(\bw_k^H\nabla f\left(\bw_k\right)\right)\bw_k.
\end{equation}

Moreover, $\mathcal{T}_{\bw_{k-1}}\left(\bm{\eta}_{k-1}\right)$ in \eqref{eta_k} denotes the vector transport of $\bm{\eta}_{k-1}$, which is defined as the projection of $\bm{\eta}_{k-1}$ to $T_{\bw_k}\mathcal{S}$ and is given as follows  
\begin{equation} \label{T_eta}
	\mathcal{T}_{\bw_{k-1}}\left(\bm{\eta}_{k-1}\right) = \bm{\eta}_{k-1} - \mathfrak{R}\left(\bw_k^H\bm{\eta}_{k-1}\right)\bw_k.
\end{equation}

Here, we choose the Polak-Ribi$\grave{\mathrm{e}}$re parameter as $\mu_k$ to achieve fast convergence, which is given by \cite{article1}
\begin{equation}	\label{muk}
	\mu_k = \frac{\langle\mathrm{grad}f\left(\bw_k\right),\mathrm{grad}f\left(\bw_k\right) - \mathcal{T}_{\bw_{k-1}}\left(\mathrm{grad}f\left(\bw_{k-1}\right)\right)\rangle}{\langle\mathrm{grad}f\left(\bw_{k-1}\right),\mathrm{grad}f\left(\bw_{k-1}\right)\rangle},
\end{equation}
where $\langle \bx,\by \rangle = \mathfrak{R}\left(\bx^H\by\right)$ and $\mathcal{T}_{\bw_{k-1}}\left(\mathrm{grad}f\left(\bw_{k-1}\right)\right)$ denotes the vector transport of $\mathrm{grad}f\left(\bw_{k-1}\right)$ which has the same form as \eqref{T_eta}. Besides, the value satisfying the Armijo condition \cite{Numericaloptimization} is chosen as the step size $s_k$, which is expressed as
\begin{equation} \label{Armijo}
	f\left(\bw_{k+1}\right) \leq f\left(\bw_k\right) + c_1 s_k\langle\mathrm{grad}f\left(\bw_k\right),\bm{\eta}_k\rangle,
\end{equation}  
where $0 < c_1 < 1$. 
%After obtaining $\bw_{k + 1}$, let $k = k + 1$ and when $\left\|\mathrm{grad} f\left(\bw_k\right)\right\|$ is less than the predetermined accuracy, it is then treated as the solution to $\mathcal{P}_{14}$.
The algorithm for solving problem $\mathcal{P}_{14}$ is summarized in Algorithm \ref{op_w}. 
\begin{algorithm}[h]
	\caption{\it Algorithm for Solving Problem $\mathcal{P}_{14}$}
	\label{op_w}
	\begin{algorithmic}
		\State 1. Initialize randomly $\bw_0 \!\in\! \mathcal{S}$, set iteration index $k \!=\! 0$, target convergence accuracy $\epsilon_2$ and maximum number of iterations $I_{max}$.
		\State 2. Set $\bm{\eta}_0 = -\mathrm{grad} f\left(\bw_0\right)$.
		\Repeat
		\State 3. Calculate $\bw_{k + 1}$ by the retraction \eqref{w_1} and choose the step size $s_k$ according to \eqref{Armijo}.
		\State 4. Compute $\mu_{k + 1}$ according to \eqref{muk}.
		\State 5. Calculate $\mathcal{T}_{\bw_{k}}\left(\bm{\eta}_{k}\right)$ by \eqref{T_eta}, and compute the search direction $\bm{\eta}_{k + 1}$ by \eqref{eta_k}.
		\State 6. Set $k = k + 1$.
		\Until $\|\mathrm{grad}f\left(\bw_{k}\right)\| \le \epsilon_2$ or $ k \geq I_{max}$.
		\State 7. Output $\bw = \bw_k$.
	\end{algorithmic}
\end{algorithm}

With $\bw$ fixed, the subproblem of optimizing $\bq$ can be formulated as
\begin{equation} \label{p8}
	\mathcal{P}_{15}:
	\min_{\bq} f\left(\bq\right) ~\mathrm{s.t.} ~ \bq \in \mathcal{O},
\end{equation}
where $f\left(\bq\right) = \frac{1 + P/\sigma_\mathrm{E}^2\left(\bq^H\bF_{\mathrm{E}}\bq + \bq^H\bm{f}_{\mathrm{E}} + \bm{f}_{\mathrm{E}}^H\bq + g_{\mathrm{E}}\right)}{1 + P/\sigma_\mathrm{B}^2\left(\bq^H\bF_{\mathrm{B}}\bq + \bq^H\bm{f}_{\mathrm{B}} + \bm{f}_{\mathrm{B}}^H\bq + g_{\mathrm{B}}\right)} + \zeta\epsilon_1\ln (1 + $  $e^{\frac{\gamma  - \gamma_{\mathrm{A}}}{\epsilon_1}})$.
%\begin{equation}
%	\notag
%	\begin{aligned}
%	f\left(\bq\right) = & \frac{1 + P/\sigma_\mathrm{E}^2\left(\bq^H\bF_{\mathrm{E}}\bq + \bq^H\bm{f}_{\mathrm{E}} + \bm{f}_{\mathrm{E}}^H\bq + g_{\mathrm{E}}\right)}{1 + P/\sigma_\mathrm{B}^2\left(\bq^H\bF_{\mathrm{B}}\bq + \bq^H\bm{f}_{\mathrm{B}} + \bm{f}_{\mathrm{B}}^H\bq + g_{\mathrm{B}}\right)}  \\
%	+ &\zeta\epsilon_1\ln\left(1 + e^{\frac{\gamma  - \gamma_{\mathrm{A}}}{\epsilon_1}}\right).
%	\end{aligned}
%\end{equation}

{{Due to the fact that it is difficult to derive the gradient of $\gamma_{\mathrm{A}}$ with respect to $\bq$, we first transform $\gamma_{\mathrm{A}}$ as follows.}} 

Note that $\gamma_{\mathrm{A}} = \frac{P\mathrm{tr}\left(\bH_{\mathrm{E}}^H\bH_{\mathrm{E}}\bW\right)}{N\sigma_{\mathrm{A}}^2} = \frac{P\|\bh_{\mathrm{E}}\|^2\mathrm{tr}\left(\bH_{\mathrm{E}}\bW\right)}{N\sigma_{\mathrm{A}}^2}$.
%\begin{equation}
%	\begin{aligned}
%		\gamma_{\mathrm{A}} = \frac{P\mathrm{tr}\left(\bH_{\mathrm{E}}^H\bH_{\mathrm{E}}\bw\bw^H\right)}{N\sigma_{\mathrm{A}}^2} = \frac{P\|\bh_{\mathrm{E}}\|^2\mathrm{tr}\left(\bH_{\mathrm{E}}\bw\bw^H\right)}{N\sigma_{\mathrm{A}}^2}.
%	\end{aligned}
%\end{equation}
The term $\|\bh_{\mathrm{E}}\|^2$ is lower bounded by 
\begin{equation} 
	\begin{aligned}
	\|\bh_{\mathrm{E}}\|^2 = & ~\bq^H\bF_{\mathrm{E}_1}\bq + \bq^H\bm{f}_{\mathrm{E}_1} + \bm{f}_{\mathrm{E}_1}^H\bq + g_{\mathrm{E}_1} \\
	\geq &  ~\bq^H\bt_1 + \bt_1^H\bq + t_1,	
	\end{aligned}
\end{equation}
where  $g_{\mathrm{E}_1} = \bh_{\mathrm{AE}}^H\bh_{\mathrm{AE}}$, $\bm{f}_{\mathrm{E}_1} = \mathrm{Diag}\left(\bh_{\mathrm{IE}}\bh_{\mathrm{AE}}^H\bH_{\mathrm{AI}}^H\right)$, $\bF_{\mathrm{E}_1} = \left(\bh_{\mathrm{IE}}\bh_{\mathrm{IE}}^H\right)\odot\left(\left(\bH_{\mathrm{AI}}\bH_{\mathrm{AI}}^H\right)^T\right)$, $\bt_1 = \bF_{\mathrm{E}_1}^H\bq_0 + \bm{f}_{\mathrm{E}_1}$, and $t_1 = g_{\mathrm{E}_1} - \bq_0^H\bF_{\mathrm{E}_1}\bq_0$ with $\bq_0$ being a feasible value of $\bq$. 
For the term $\mathrm{tr}\left(\bH_{\mathrm{E}}\bW\right)$, it can be lower bounded as follows
\begin{equation}
	\begin{aligned}
		\mathrm{tr}\left(\bH_{\mathrm{E}}\bW^H\right) = & ~\bq^H\bF_{\mathrm{E}}\bq + \bq^H\bm{f}_{\mathrm{E}} + \bm{f}_{\mathrm{E}}^H\bq + g_{\mathrm{E}} \\
		\geq &  ~\bq^H\bt_2 + \bt_2^H\bq + t_2,	
	\end{aligned}
\end{equation}
where $\bt_2 = \bF_{\mathrm{E}}^H\bq_0 + \bm{f}_{\mathrm{E}}$, and $t_2 = g_{\mathrm{E}} - \bq_0^H\bF_{\mathrm{E}}\bq_0$.
Thus, $\gamma_{\mathrm{A}}$ can be approximated by
\begin{equation} \label{gamma_A'}
	\notag
	\begin{aligned}
		\gamma_{\mathrm{A}} \geq \tilde{\gamma}_{\mathrm{A}} 
		=   \frac{P\left(\bq^H\bt_1 + \bt_1^H\bq + t_1\right)\left(\bq^H\bt_2 + \bt_2^H\bq + t_2\right)}{N\sigma_{\mathrm{A}}^2}.
	\end{aligned}
\end{equation}

Based on the above approximation, the penalty function can be transformed into $p\left(\bq\right) = \zeta\epsilon_1\ln\left(1 + e^{\frac{\gamma  - \tilde{\gamma}_{\mathrm{A}}\left(\bq\right)}{\epsilon_1}}\right)$.
%\begin{equation}
%	p\left(\bq\right) = \zeta\epsilon_1\ln\left(1 + e^{\frac{\gamma  - \tilde{\gamma}_{\mathrm{A}}\left(\bq\right)}{\epsilon_1}}\right).
%\end{equation}

Further, the subproblem of $\bq$ can be expressed as
\begin{equation} \label{p9}
	\notag
	\mathcal{P}_{16}:
	\min_{\bq} f\left(\bq\right) ~\mathrm{s.t.} ~ \bq \in \mathcal{O},
\end{equation}
where $f\left(\bq\right) =  \frac{1 + P/\sigma_\mathrm{E}^2\left(\bq^H\bF_{\mathrm{E}}\bq + \bq^H\bm{f}_{\mathrm{E}} + \bm{f}_{\mathrm{E}}^H\bq + g_{\mathrm{E}}\right)}{1 + P/\sigma_\mathrm{B}^2\left(\bq^H\bF_{\mathrm{B}}\bq + \bq^H\bm{f}_{\mathrm{B}} + \bm{f}_{\mathrm{B}}^H\bq + g_{\mathrm{B}}\right)} + p\left(\bq\right)$.
%\begin{equation}
%	\notag
%	\begin{aligned}
%		f\left(\bq\right) = & \frac{1 + P/\sigma_\mathrm{E}^2\left(\bq^H\bF_{\mathrm{E}}\bq + \bq^H\bm{f}_{\mathrm{E}} + \bm{f}_{\mathrm{E}}^H\bq + g_{\mathrm{E}}\right)}{1 + P/\sigma_\mathrm{B}^2\left(\bq^H\bF_{\mathrm{B}}\bq + \bq^H\bm{f}_{\mathrm{B}} + \bm{f}_{\mathrm{B}}^H\bq + g_{\mathrm{B}}\right)} + p\left(\bq\right).
%	\end{aligned}
%\end{equation}

Next, the RCG algorithm is adopted. 
Given the $k$-th iteration point $\bq_k$, the next point $\bq_{k + 1}$ can be calculated as
\begin{equation}
	\begin{aligned}	\label{q1}
		\bq_{k + 1}  = R_{\bq_k}\left(s_k\bm{\eta}_k\right) 
		 = \mathrm{unt}\left(\bq_k + s_k\bm{\eta}_k\right),	
	\end{aligned}
\end{equation}
where $\mathrm{unt}\left(\bx\right) = \left[\frac{x_1}{\left|x_1\right|},\frac{x_2}{\left|x_2\right|},\dots,\frac{x_M}{\left|x_M\right|}\right]^T$. Since the process of calculating $\bm{\eta}_k$ is similar to that of $\bw$, we provide the corresponding formulas directly as follows.

For oblique manifold $\mathcal{O}$, the tangent space at point $\bq_k$ is $T_{\bq_k}\mathcal{O} \!=\! \left\{\bv \in \mathbb{C}^{M \times 1} \left| \left[\bv\bq_k^H\right]_{m,m} \!=\! 0,m = 1,\dots,M \right. \right\}$.
%\begin{equation} \label{TqO}
%	T_{\bq_k}\mathcal{O} \!=\! \left\{\bv \in \mathbb{C}^{M \times 1} \left| \left[\bv\bq_k^H\right]_{m,m} \!=\! 0,m = 1,\dots,M \right. \right\}.
%\end{equation}
The Euclidean gradient of $f\left(\bq\right)$ at point $\bq_k$ is calculated as
\begin{equation}
	\begin{aligned}		\label{nablaq}
		\nabla f\left(\bq_k\right) = &
		2P\left\{\frac{\bF_{\mathrm{E}}\bq_k + \bm{f}_{\mathrm{E}}}{\sigma_\mathrm{E}^2 g_{\mathrm{bk}}} - 
		\frac{\left(\bF_{\mathrm{B}}\bq_k + \bm{f}_{\mathrm{B}}\right)g_{\mathrm{ek}}}{\sigma_\mathrm{B}^2 g_{\mathrm{bk}}^2}  \right.\\
		- & \left. \frac{\zeta \left(t_{2_k}\bt_1 + t_{1_k}\bt_2\right)}{N\sigma_{\mathrm{A}}^2 \left(1 \!+\! e^{\frac{\tilde{\gamma}_{\mathrm{A}}\left(\bq_k\right) - \gamma}{\epsilon_1}}\right)}\right\},
	\end{aligned}
\end{equation} 
where $g_{\mathrm{ek}} = 1 \!+\! P/\sigma_\mathrm{E}^2\left(\bq_k^H\bF_{\mathrm{E}}\bq_k + \bq_k^H\bm{f}_{\mathrm{E}} + \bm{f}_{\mathrm{E}}^H\bq_k + g_{\mathrm{E}}\right)$, $g_{\mathrm{bk}} = 1 \!+\! P/\sigma_\mathrm{B}^2\left(\bq_k^H\bF_{\mathrm{B}}\bq_k + \bq_k^H\bm{f}_{\mathrm{B}} + \bm{f}_{\mathrm{B}}^H\bq_k + g_{\mathrm{B}}\right)$, $t_{1_k} = \bq_k^H\bt_1 + \bt_1^H\bq_k + t_1$, and $t_{2_k} = \bq_k^H\bt_2 + \bt_2^H\bq_k + t_2$. 

The Riemannian gradient of $f\left(\bq\right)$ at point $\bq_k$ is expressed as $\mathrm{grad}f\left(\bq_k\right) = \nabla f\left(\bq_k\right) - \mathfrak{R}\{\nabla f\left(\bq_k\right)\odot\bq_k^*\}\odot\bq_k$ \cite{7397861}.
%\begin{equation}
%	\mathrm{grad}f\left(\bq_k\right) = \nabla f\left(\bq_k\right) - \mathfrak{R}\{\nabla f\left(\bq_k\right)\odot\bq_k^*\}\odot\bq_k.
%\end{equation} 
Moreover, the vector transport is defined as
\begin{equation} \label{T_etaq}
	\mathcal{T}_{\bq_{k-1}}\left(\bm{\eta}_{k-1}\right) = \bm{\eta}_{k-1} - \mathfrak{R}\left\{\bm{\eta}_{k-1}\odot\bq_k^*\right\}\odot\bq_k.
\end{equation} 

The Polak-Ribi$\grave{\mathrm{e}}$re parameter is still chosen as $\mu_k$ which can be obtained by substituting $\bq$ for $\bw$ in \eqref{muk} and $s_k$ satisfying the Armijo condition is chosen as the step size. 
%After obtaining $\bq$, the phase shift matrix can be set as $\bQ = \mathrm{diag}\left(\bq\right)$. 
The algorithm of optimizing $\bq$ is similar to Algorithm \ref{op_w} and is omitted here. We summarize the whole RCG-based AO algorithm for solving problem $\mathcal{P}_{13}$ in Algorithm \ref{rcg}.

 \begin{algorithm}[htbp]
	\caption{\it RCG-based AO Algorithm for Solving $\mathcal{P}_{13}$}
	\label{rcg}
	\begin{algorithmic} 
		\State 1. Set initial points $\bw_0 \in \mathcal{S}$ and $\bq_0 \in \mathcal{O}$, $\bQ_0 = \mathrm{diag}\left(\bq_0\right)$, iteration index $i = 0$, a proper penalty factor $\zeta$, $\epsilon_1$, and convergence accuracy $\epsilon_3$.
		\State 2. Calculate $f_0 = f\left(\bw_0, \bQ_0\right)$ according to \eqref{p6}.
		\Repeat
		\State 3. Set $ i = i + 1$
		\State 4. Obtain $\bw_i$ for given $\bq_{i - 1}$ by Algorithm \ref{op_w}.
		\State 5. Obtain $\bq_i$ for given $\bw_i$ by RCG algorithm and set $\bQ_i = \mathrm{diag}\left(\bq_i\right)$. 
		\State 6. Calculate $f_i = f\left(\bw_i, \bQ_i\right)$.   
		\Until 
		$\frac{\left|f_i -f_{i - 1}\right|}{\left|f_{i - 1}\right|}\leq \epsilon_3$.
		\State 7. Output $\bw_i$ and $\bQ_i$.
	\end{algorithmic}
\end{algorithm}	

\emph{Complexity Analysis:} For the Dinkelbach method-based AO algorithm, the main complexity arises from solving the SDP problems for $\bW$ and $\bU$. For the SDP problems of $\bW$ and $\bU$, we have $m \!\!\!=\!\!\! 2, n \!\!\!=\!\!\! N$ and $m \!\!=\!\! M \!+\! 2, n \!\!=\!\! M \!+\! 1$, respectively. Hence, the total complexity is about $\mathcal{O}\left(2N^3 + 4N^2 + 8  + \left(M + 2\right)\left(M + 1\right)^3 \!+\! \left(M + 2\right)^2\left(M + \right. \right.$ $\left.  \!\left.1\right)^2  +\! \left(M + 2\right)^3\right)$ in each iteration. As for the RCG-based algorithm, the main computational complexity comes from the retraction, Euclidean gradient, Riemannian gradient, and the vector transport in each iteration, which is about $\mathcal{O}\left(N^2 + N\right)$ for $\bw$ and $\mathcal{O}\left(M^2 + M\right)$ for $\bQ$ \cite{2004Convex}. Thus, the overall complexity of Algorithm \ref{rcg} is about $\mathcal{O}\left(N^2 + N + M^2 + M\right)$ in each iteration.

\section{Robust secure communication in DFRC system}
The results obtained in previous section, assuming perfect CSI, serve as a fundamental secure communication scheme for DFRC system. However, in practice, targets typically remain passive who do not interact with the BS actively, making it challenging for BS to acquire their accurate CSI \cite{9199556}. {{Consequently, we extend our investigation to consider the impact of target's imperfect CSI on RIS-aided secure communication in DFRC system}}\footnote{{{We note that the following design framework can be extended to the RCCE system with target's imperfect CSI.}}}. For simplicity, the radar illumination power at the target is adopted as the performance metric of radar detection \cite{9647914} in this section.
%based on the assumption that the received SINR at BS increases with the radar illumination power at target. 
%We first employ a bounded uncertainty model to capture the angle error and fading channel error of the eavesdropper, and derive a tractable bound for their joint uncertainty. Then, a secrecy rate maximization problem under eavesdropper's imperfect CSI is formulated by jointly designing the transmit beamforming at the BS and the phase shifts of RIS. A BCD-based algorithm combined with S-procedure is utilized to obtain a suboptimal solution to the formulated problem. 

\subsection{Problem Formulation} 
Recalling the channel modeling in Section II, a bounded uncertainty model is adopted to characterize the estimation error of cascaded channel Alice-RIS-Eve which is given by 
\begin{equation}
	\begin{aligned}
		\bG & = \bar{\bG} + \Delta\bG,	\\
		\Omega_\mathrm{G} & = \left\{\Delta\bG \in \mathbb{C}^{M \times N}:\Vert\Delta\bG\Vert_F \leq \epsilon_\mathrm{G}\right\},
	\end{aligned} 
\end{equation}     
where $\bG = \mathrm{diag}\left(\bh_{\mathrm{IE}}^H\right)\bH_{\mathrm{AI}}$, $\bar{\bG}$ is the estimate of the cascaded channel; $\Delta\bG$ is the estimate uncertainty bounded by $\epsilon_\mathrm{G}$. We note that the bounded uncertainty model is widely-adopted to account for practical estimation errors \cite{9180053,9658554}. Moreover, the angle estimation error is considered for channel $\bh_{\mathrm{AE}}$ \cite{9199556}. In specific, the angle $\theta$ is modeled as $\theta \!\!=\!\! \bar{\theta} \!\!+\!\! \Delta\theta$ where $\bar{\theta}$ is the estimate of target's angle and $\Delta\theta$ is the angle uncertainty which is bounded by $\phi$, i.e., $\left|\Delta\theta\right| \leq \phi$. {{The bounded uncertainty $\Delta\bG$ can be tackled by employing suitable mathematical transformations. However, it is challenging to deal with angle uncertainty. Next, a tractable bound characterizing the effect of angle uncertainty is derived.}} 

Via the safe approximation \cite{9933849}, $\ba_{\mathrm{A}}$ in \eqref{a_A} can be recast as
\begin{equation}
	\ba_{\mathrm{A}} = \bar{\ba}_A + \Delta\ba_{\mathrm{A}},
\end{equation}
where $\bar{\ba}_A = \left[1,e^{j\frac{2\pi d}{\lambda}\sin\bar{\theta}},\right.$ $\left.\dots,e^{j\frac{2\pi d}{\lambda}\left(N - 1\right)\sin\bar{\theta}}\right]^T$ can be regarded as the estimate of $\ba_{\mathrm{A}}$, and $\Delta\ba_{\mathrm{A}} = \left[a_1,a_2,\dots,a_N\right]^T$. Each entry in $\Delta\ba_{\mathrm{A}}$ is bounded as follows
\begin{equation}
	\left|a_n\right| \leq \sqrt{2\left(1 - \cos\psi_n\right)}, n\in\{1,\dots,N\},
\end{equation}
where $\psi_n = \left|\frac{2\pi d}{\lambda}\left(n - 1\right)\left(\sin \bar{\theta} - \sin\left(\bar{\theta} + \phi\right)\right)\right|$
Thus, we can obtain a bound of $\Delta\ba_{\mathrm{A}}$ as follows
\begin{equation}
	\Vert\Delta\ba_A\Vert \leq \epsilon_{\mathrm{A}}, 
\end{equation} 
where $\epsilon_{\mathrm{A}} = \sqrt{\sum_{n = 1}^{N}2\left(1 - \cos\psi_n\right)}$.  
Furthermore, we can rewrite $\bh_{\mathrm{AE}}$ as
\begin{equation}
	\bh_{\mathrm{AE}} = \bar{\bh}_{\mathrm{AE}} + \Delta\bh_{\mathrm{AE}},
\end{equation}
where $\bar{\bh}_{\mathrm{AE}} = \beta_{\mathrm{AE}}\bar{\ba}_A$, and $\Delta\bh_{\mathrm{AE}} = \beta_{\mathrm{AE}}\Delta\ba_{\mathrm{A}}$ is the estimation error constrained by $\Vert\Delta\bh_{\mathrm{AE}}\Vert \leq \epsilon_{\mathrm{AE}}$ with $\epsilon_{\mathrm{AE}} = \beta_{\mathrm{AE}}\epsilon_{\mathrm{A}}$. 

Now, the secrecy rate maximization problem with target's imperfect CSI is formulated as follows
	\begin{subequations}
	\begin{align}
		\label{POF}
		\mathcal{P}^{''}:
		\max_{\bw,\bq}~ 
		&\left(C_\mathrm{B} - \max_{\Delta\bh_{\mathrm{AE}},\Delta\bG}C_\mathrm{E}\right) \\
		\mathrm{s.t.}~& \Vert\bw\Vert = 1, \\
		\label{consq}
		& \left|q_i\right| = 1, ~\forall i \in \{1, \dots, M\}, \\
		\label{consr}
		& \left\|y_{\mathrm{E}}\right\|^2 \geq \gamma_p, ~\forall \Delta\bh_{\mathrm{AE}},\Delta\bG,
	\end{align}
\end{subequations}
where $q_i$ is the $i$-th entry of $\bq$. Here, \eqref{POF} indicates that we consider a worst-case secure communication design where the BS always encounters the worst estimate of target's CSI. \eqref{consr} specifies the radar detection constraint which requires that the radar illumination power at the target should be no less than a predetermined threshold $\gamma_p$.

{{We note that problem $\mathcal{P}^{''}$ is non-convex. The non-convexity stems from the highly coupled variables in both the objective function and constraint \eqref{consr}, and the typically non-convex constraint \eqref{consq}. Moreover, the CSI estimation error also hinders the design of an efficient algorithmic solution to the problem. Next, we first transform the problem into a tractable form and adopt a BCD-based algorithm to tackle it.}}

\subsection{Problem Solution}
Define $\bW \triangleq \bw\bw^H$, $\boldsymbol{\Theta} \triangleq \tilde{\bq}^H\tilde{\bq}$ with $\tilde{\bq} = \left[\bq^T,1\right]$, $\bB = \left[\begin{array}{c}
	\mathrm{diag}\left(\bh_{\mathrm{IB}}^H\right)\bH_{\mathrm{AI}}	\\
	\bh_{\mathrm{AB}}^H
\end{array}\right]$, and $\bE = \left[\begin{array}{c}
	\bG	\\
	\bh_{\mathrm{AE}}^H
\end{array}\right]$. Also, $\bE$ can be rewritten as follow
\begin{equation} \label{E}
	\bE = \bar{\bE} + \Delta\bE = \left[\begin{array}{c}
		\bar{\bG}	\\
		\bar{\bh}_{\mathrm{AE}}^H
	\end{array}\right] + \left[\begin{array}{c}
		\Delta\bG	\\
		\Delta\bh_{\mathrm{AE}}^H
	\end{array}\right].
\end{equation}
Via some mathematical manipulations, we can derive that $\Vert\Delta\bE\Vert_F \leq \epsilon_E$ with $\epsilon_E = \sqrt{\epsilon_{\mathrm{G}}^2 + \epsilon_{\mathrm{AE}}^2}$. Then, the optimization problem can be reformulated as follows
\begin{subequations}
	\begin{align}
		\mathcal{P}^{''}_1:
		~\max_{\bW,\boldsymbol{\Theta}}~&\left(C_\mathrm{B} - \max_{\Delta\bE}C_\mathrm{E}\right) \\
		\label{p1w}
		\mathrm{s.t.}~& \bW = \bW^H, \bW \succeq \bf{0},\mathrm{tr}\left(\bW\right) = 1,\\
		\label{p1wr}
		& \mathrm{rank}\left(\bW\right) = 1,\\
		\label{p1q}
		& \boldsymbol{\Theta} = \boldsymbol{\Theta}^H, \boldsymbol{\Theta} \succeq \boldsymbol{0}, \boldsymbol{\Theta}_{i,i} = 1, \forall i \in \mathcal{M}_1,\\
		\label{p1qr}
		& \mathrm{rank}\left(\boldsymbol{\Theta}\right) = 1, \\
		\label{p1r}
		& \mathrm{tr}\left(\boldsymbol{\Theta}\bE\bW\bE^H\right) \geq \gamma_p/P, \forall \Delta\bE,
	\end{align}
\end{subequations} 
where $C_\mathrm{J} \!=\! \log_2\left(1 + P/\sigma_\mathrm{J}^2\mathrm{tr}\left(\boldsymbol{\Theta}\bJ\bW\bJ^H\right)\right)$ with $\mathrm{J} \in \left\{\mathrm{B}, \mathrm{E}\right\}$, $\bJ \in \left\{\bB, \bE\right\}$ and $\mathcal{M}_1 = \left\{1,\dots, M + 1\right\}$.

To make the second term in the objective function tractable, we introduce two slack variables $\eta$ and $\kappa$, and rewrite the problem as follows
\begin{subequations}
	\begin{align}
		\mathcal{P}^{''}_2:
		~\max_{\bW,\boldsymbol{\Theta},\eta,\kappa}~&\left(C_\mathrm{B} - \eta\right) \\
		\label{eta}
		\mathrm{s.t.}~& \eta \geq \log_2\left(1 + \kappa\right),\\
		\label{kappa}
		& \kappa \sigma_\mathrm{E}^2/P \geq \max_{\Delta\bE} \mathrm{tr}\left(\boldsymbol{\Theta}\bE\bW\bE^H\right), \\
		\notag
		& \eqref{p1w}-\eqref{p1r}.
	\end{align}
\end{subequations}

% Till now, the optimization problem becomes more tractable, but there are still some difficulties, such as the semi-infinite constraints \eqref{p1r} and \eqref{kappa} caused by the channel uncertainty, and the non-convex constraint \eqref{eta}. 
Next, the S-procedure is utilized to handle the semi-infinite constraints \eqref{p1r} and \eqref{kappa}, which transforms them into their equivalent linear matrix inequalities. Interested readers can refer to \cite{2004Convex} for more details of the S-procedure.
%We first briefly introduce the S-procedure \cite{2004Convex}.
%\emph{S-procedure}: Let $f_k\left(\bx\right), k = \left\{1,2\right\}$, be
%\begin{equation}
%	f_k\left(\bx\right) = \bx^H\bA_k\bx + 2\Re\left\{\bb_k^H\bx\right\} + c_k,
%\end{equation} 
%where $\bA_k$ is an Hermitian matrix, $\bx$ and $\bb_k$ are complex vectors, and $c_k$ is a real scalar. Assume that there exists $\hat{\bx}$ such that $f_1\left(\hat{\bx}\right) < 0$. Then, the implication $f_1\left(\bx\right) \leq 0 \Rightarrow f_2\left(\bx\right) \leq 0$ holds true if and only if there exists $\rho \geq 0$ such that
%\begin{equation}
%	\rho\left[\begin{array}{cc}
%		\bA_1 & \bb_1 \\
%		\bb_1^H & c_1
%	\end{array}\right] - \left[\begin{array}{cc}
%		\bA_2 & \bb_2 \\
%		\bb_2^H & c_2
%	\end{array}\right] \succeq \boldsymbol{0}.
%\end{equation}

To facilitate the S-procedure, we need to transform constraints \eqref{p1r} and \eqref{kappa} as follows. First, the uncertainty constraint $\Vert\Delta\bE\Vert_F \leq \epsilon_E$ can be equivalently written as 
\begin{equation}
	\be^H\be -\epsilon_E^2 \leq 0, \be = \mathrm{vec}\left(\Delta\bE\right).
\end{equation}

Based on \eqref{E}, constraint \eqref{p1r} can be further expressed as 
\begin{equation} \label{ts1}
	\begin{aligned}
		&\mathrm{tr}\left(\boldsymbol{\Theta}\bE\bW\bE^H\right) \geq \gamma_p \\
		& \Leftrightarrow \mathrm{tr}\left(\boldsymbol{\Theta}\bar{\bE}\bW\bar{\bE}^H + \boldsymbol{\Theta}\bar{\bE}\bW\Delta\bE^H + \boldsymbol{\Theta}\Delta\bE\bW\bar{\bE}^H \right. \\
		& \left. + ~\boldsymbol{\Theta}\Delta\bE\bW\Delta\bE^H\right) \geq \gamma_p.
	\end{aligned}
\end{equation}

Using $\mathrm{tr}\left(\bX\bY\bZ\bU\right) = \left(\mathrm{vec}^H\left(\bU^H\right)\right)\left(\bZ^H\otimes\bX\right)\mathrm{vec}\left(\bY\right)$, we can have $\mathrm{tr}\left(\boldsymbol{\Theta}\Delta\bE\bW\Delta\bE^H\right) = \be^H\left(\bW\otimes\boldsymbol{\Theta}\right)\be$. Moreover, according to $\mathrm{tr}\left(\bX^H\bY\right) = \mathrm{vec}^H\left(\bX\right)\mathrm{vec}\left(\bY\right)$, we can have $\mathrm{tr}\left(\boldsymbol{\Theta}\Delta\bE\bW\bar{\bE}^H\right) = \bp^H\be$ where $\bp = \mathrm{vec}\left(\boldsymbol{\Theta}\bar{\bE}\bW\right)$. Thus, the inequality in \eqref{ts1} can be further written as 
\begin{equation}
	\be^H\left(\bW\otimes\boldsymbol{\Theta}\right)\be + 2\Re\left\{\bp^H\be\right\} + c_E \geq \gamma_p,
\end{equation} 
where $c_E = \mathrm{tr}\left(\boldsymbol{\Theta}\bar{\bE}\bW\bar{\bE}^H\right)$.

Now, using S-procedure, constraint \eqref{p1r} can be recast as
\begin{equation}	\label{semi1}
	\left[\begin{array}{cc}
		\rho_1\bI + \bW\otimes\boldsymbol{\Theta} & \bp \\
		\bp^H & c_E - \gamma_p/P - \rho_1\epsilon_E^2
	\end{array}\right] \succeq \boldsymbol{0}.
\end{equation}
Similarly, constraint \eqref{kappa} can be recast as
\begin{equation} \label{semi2}
	\left[\begin{array}{cc}
		\rho_2\bI - \bW\otimes\boldsymbol{\Theta} & -\bp \\
		-\bp^H &  \kappa\sigma_\mathrm{E}^2/P -c_E - \rho_2\epsilon_E^2
	\end{array}\right] \succeq \boldsymbol{0}.	
\end{equation}

For the non-convex constraint \eqref{eta}, a convex lower bound of it is derived based on the first-order Taylor expansion as
\begin{equation} \label{tseta}
	\eta \geq \log_2\left(1 + \kappa\right) \geq \frac{1}{\mathrm{ln}\left(2\right)}\frac{\kappa - \kappa_0}{\left(1 + \kappa_0\right)} + \log_2\left(1 + \kappa_0\right),
\end{equation}  
where $\kappa_0$ denotes a feasible value of $\kappa$.

Finally, the problem $\mathcal{P}^{''}_2$ is transformed into
\begin{subequations}
	\begin{align}
		\notag
		\mathcal{P}^{''}_3:
		~\max_{\bW,\boldsymbol{\Theta},\eta,\kappa}~&\left(C_\mathrm{B} - \eta\right) \\
		\notag
		\mathrm{s.t.}~& \eqref{p1w}-\eqref{p1qr}, \eqref{semi1}-\eqref{tseta}.
	\end{align}
\end{subequations}

{{A BCD-based algorithm is exploited to obtain a suboptimal solution to it.}} Specifically, we decompose $\mathcal{P}^{''}_3$ into the following two subproblems
\begin{subequations}
	\begin{align}
		\notag
		\mathcal{P}^{''}_4:
		~\max_{\bW,\eta,\kappa}~\left(C_\mathrm{B} - \eta\right) ~
		\mathrm{s.t.}~ \eqref{p1w},\eqref{p1wr}, \eqref{semi1}-\eqref{tseta}.
	\end{align}
\end{subequations} 
\begin{subequations}
	\begin{align}
		\notag
		\mathcal{P}^{''}_5:
		~\max_{\boldsymbol{\Theta},\eta,\kappa}~\left(C_\mathrm{B} - \eta\right)~
		\mathrm{s.t.}~ \eqref{p1q},\eqref{p1qr}, \eqref{semi1}-\eqref{tseta}.
	\end{align}
\end{subequations}

The specific process and the computational complexity analysis for solving these problems are similar to that in Section II and are omitted for brevity. The whole algorithm for solving problem $\mathcal{P}^{''}$ is summarized in Algorithm \ref{BCD Imperfect}.  

\begin{algorithm}[h]
	\caption{\it BCD-based Algorithm for Solving $\mathcal{P}^{''}$}
	\label{BCD Imperfect}
	\begin{algorithmic}
		\State 1. Set iteration index $k = 0$. Initialize randomly $\bw_k$ and $\bq_k$, and set $\bW_k = \bw_k\bw_k^H$, $\boldsymbol{\Theta}_k = \tilde{\bq}_k^H\tilde{\bq}_k$, $\kappa_k = 0$, $C_{s_k} = 0$.
		\State 2. Set convergence accuracy $\varepsilon$.
		\Repeat
		\State 3. Obtain $\bW_{k + 1}$ by solving problem $\mathcal{P}^{''}_4$ for given $\boldsymbol{\Theta}_k$.
		\State 4. Obtain $\boldsymbol{\Theta}_{k + 1}$ and $\eta_{k + 1}$ by solving problem $\mathcal{P}^{''}_5$ for given $\bW_{k + 1}$.
		\State 5. Calculate $C_{s_{k + 1}} = C_\mathrm{B}\left(\bW_{k + 1},\boldsymbol{\Theta}_{k + 1}\right) - \eta_{k + 1}$.
		\State 6. Set $k = k + 1$.
		\Until $ \left|\frac{C_{s_{k}} - C_{s_{k - 1}}}{C_{s_{k - 1}}}\right|\leq \varepsilon$.
		\State 7. Output $\bw_k$ and $\bq_k$.
	\end{algorithmic}
\end{algorithm}

\section{Simulation results}
In this section, simulation results are presented to examine the performance of the proposed algorithms. In a three-dimensional Cartesian coordinate system, Alice, RIS, Bob, and Eve are located at $\left(0,0,0\right)$, $\left(-2.5,2.5\sqrt{3},5\right)$, $\left(0,30,10\sqrt{3}\right)$, and $\left(45,15\sqrt{3},30\right)$, respectively. Alice is equipped with a uniform linear array located at the $z$ axis. The RIS is fabricated as a square grid and is parallel to the $y-z$ plane. Unless otherwise stated, in all simulations, the path loss is modeled as $\beta_{x}^2 = \beta_0^2\left(\frac{d_x}{d_0}\right)^{-\alpha_x}$, where $\beta_0^2 = -15$ dB denotes the path loss at the reference distance $d_0 = 1$ m, $d_x$ and $\alpha_x$ denote the distance between two nodes and the path loss exponent. We set $\alpha_{\mathrm{AI}} = \alpha_{\mathrm{IB}} = \alpha_{\mathrm{IE}} = 2$, $\alpha_{\mathrm{AE}} = 2.2$, $\alpha_{\mathrm{AB}} = 3.2$. Other parameters are given in Table \ref{Simulation parameters}. 
%$N = 4$, $M = 9$, $P = 30$ dBm, $\sigma_\mathrm{B}^2 = \sigma_\mathrm{E}^2 = -60$ dBm, $\sigma_{\mathrm{A}}^2 = -110$ dBm, $\gamma = 15$ dB, $\epsilon_1 = 0.3$, $\zeta = 0.2$, $\epsilon_2 = 10^{-4}$, and $\epsilon_b = \epsilon_3 = 10^{-3}$. 
Simulation results in Fig. \ref{fig_conver} is obtained based on a randomly generated channel, and results in other figures are averaged over 100 randomly generated channels. 

\begin{table}[htbp]
	\caption{Simulation Parameters}
	\centering
	\begin{tabular}{ |c|c|c| } 
		\hline
		$N$ & Number of transmit antennas & 4 \\
		\hline 
		$M$ & Number of RIS reflecting elements & 9 \\
		\hline 
		$P$ & Total transmit power & 30 dBm \\ 
		\hline
		$\sigma_{B}^2$ & Noise power at Bob & -60 dBm \\ 
		\hline
		$\sigma_{E}^2$ & Noise power at Eve & -60 dBm \\ 
		\hline
		$\sigma_{A}^2$ & Noise power at Alice & -110 dBm \\ 
		\hline
		$\gamma$ & Received SINR threshold of the BS & 15 dB \\ 
		\hline
		$\epsilon_b$ & Convergence accuracy for Algorithm \ref{BCD} & $10^{-3}$ \\
		\hline
		$\epsilon_1$ & Given number in eq. \eqref{max_approximate} & 0.3 \\
		\hline
		$\zeta$ & Penalty factor for RCG-based AO algorithm & 0.2 \\
		\hline
		$\epsilon_2$ & Convergence accuracy for Algorithm \ref{op_w} & $10^{-4}$ \\
		\hline
		$\epsilon_3$ &Convergence accuracy for Algorithm \ref{rcg} & $10^{-3}$ \\
		\hline
		$ \varepsilon$ & Convergence accuracy for Algorithm \ref{BCD Imperfect} & $10^{-2}$ \\
		\hline
		$\bar{\theta}$ & Estimate of target angle & $30^{\circ}$ \\
		\hline
		$\phi$ & Uncertainty of target angle estimation & $3^{\circ}$ \\
		\hline
	\end{tabular}
	\label{Simulation parameters}
\end{table}

\subsection{RIS-aided Secure Communication with Perfect CSI}

In this subsection, the proposed algorithms for secure communication with perfect CSI in Sections II and III are evaluated via simulations.
         
% {convergence}
\begin{figure}[htbp]%[htbp]
	\centerline{\includegraphics[scale=0.62]{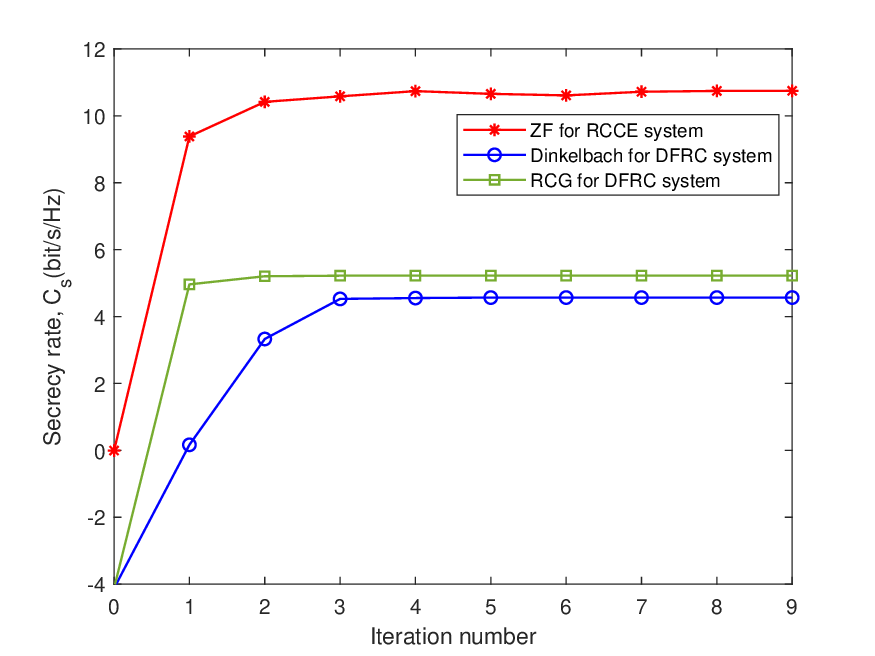}}
	\caption{Average secrecy rate $C_s$ v.s. iteration number.} \label{fig_conver}
\end{figure}
{{Fig. \ref{fig_conver} illustrates the rapid convergence of the proposed algorithms: the ZF-based BCD algorithm for RCCE system, the Dinkelbach method-based AO algorithm, and the RCG-based AO algorithm for DFRC system (labeled as ZF for RCCE system, Dinkelbach for DFRC system and RCG for DFRC system, respectively).}} From the figure, it is evident that the secrecy rates in all schemes exhibit a monotonically increasing trend with the iteration number and converge within no more than 5 iterations, which confirms the rapid convergence of the proposed algorithms. 
% Compared to the Dinkelbach method based algorithm, the RCG based algorithm converges faster.  
\begin{figure}[htbp]%[htbp]
	\centerline{\includegraphics[scale=0.62]{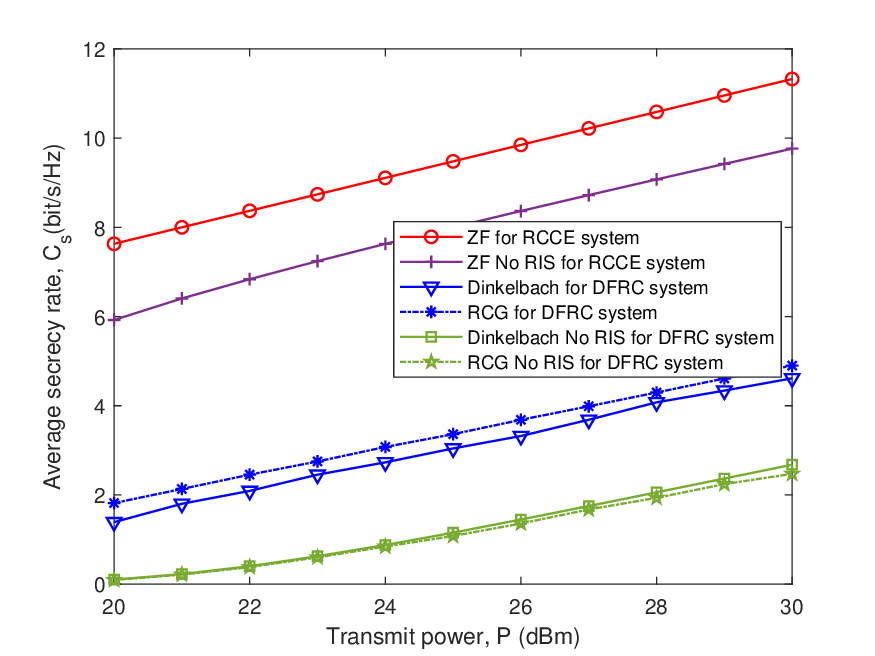}}
	\caption{Average secrecy rate $C_s$ v.s. transmit power $P$.} \label{fig_p}
\end{figure}

In Fig. \ref{fig_p}, the secrecy rate is plotted against the transmit power for various schemes, including the three proposed algorithms and their corresponding benchmark schemes without RIS. {{It can be observed that across all schemes, within the considered range of transmit power, the secrecy rate increases with the transmit power. With fixed radar detection threshold, after allocating a certain amount of power to the target to meet the radar sensing constraint, more power will be allocated to Bob. Thus, the secrecy rate continues to increase with the transmit power.}} In both considered systems, RIS leads to higher secrecy rates compared to the scenarios without RIS. Deploying RIS enables configurable signals propagation environment and introduces additional degrees of freedom, thereby boosting the secrecy performance with maintaining radar sensing performance. For the DFRC system, the RCG-based algorithm achieves slightly higher secrecy rate than the Dinkelbach method-based algorithm.
%since when solving the subproblem of $\bQ$ in the latter algorithm, a lower bound of the OF is optimized. 
Furthermore, comparing both systems, it is apparent that the RCCE system outperforms the DFRC system in terms of secrecy rate. Note that in RCCE system, ZF beamforming is employed at Alice, preventing communication signals from being received by target. Additionally, the radar signals are directed towards the nullspace of the effective channel of Alice-Bob, ensuring that radar signals won't interfere with communication. In contrast, in DFRC system, information-bearing signals are used for radar sensing concurrently, necessitating the illumination power to target to surpass a threshold so as to satisfy the radar sensing constraint, inevitably leading to the information leakage to target. Hence, higher secrecy rate is achieved in RCCE system.  

\begin{figure} [htbp]
	\centerline{\includegraphics[scale=0.62]{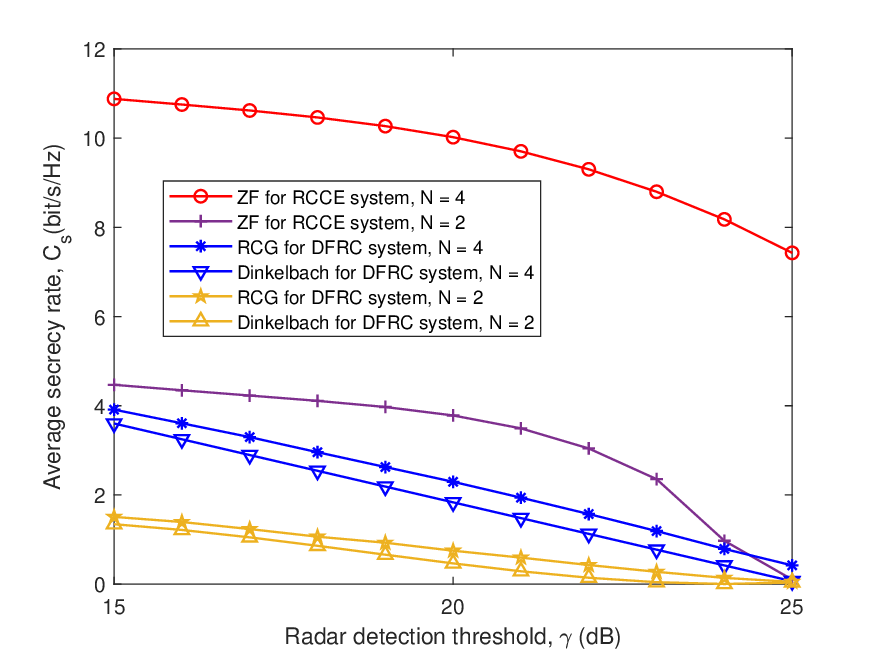}}
	\caption{Average secrecy rate $C_s$ v.s. radar detection threshold $\gamma$ with $\alpha_{\mathrm{AE}} = 2.4$.} \label{fig_gamma}
\end{figure}

Fig. \ref{fig_gamma} demonstrates the impact of radar detection threshold on secrecy rate under different numbers of transmit antennas. It is easily observed that for all schemes, the secrecy rate decreases with the radar detection threshold which illustrates the trade-off between radar sensing and secure communication that the radar sensing performance is improved at the cost of secrecy performance. 
Furthermore, when $\gamma$ is exceedingly large, such as when $N =2, \gamma = 25$ dB, the radar detection constraint cannot be satisfied, rendering the optimization problem infeasible and resulting in a zero secrecy rate. 
For RCCE system, as the radar detection threshold grows, more power is allocated to radar detection to satisfy more stringent radar sensing requirement, thereby reducing the secrecy rate. As for DFRC system, a larger $\gamma$ implies that the transmitter needs to steer the information-carrying signal towards target, leading to a lower secrecy rate. Besides, increasing the number of transmit antennas $N$ improves the secrecy rate for both systems, particularly for RCCE system. This is because as $N$ grows, more spatial degrees of freedom are available to nullify the wiretap channel and eliminate the interference introduced by radar sensing in communication. Thus, we can employ more antennas at Alice to compensate for the performance degradation caused by stricter radar detection constraint. 

\begin{figure}[htbp]	
	\centerline{\includegraphics[scale=0.62]{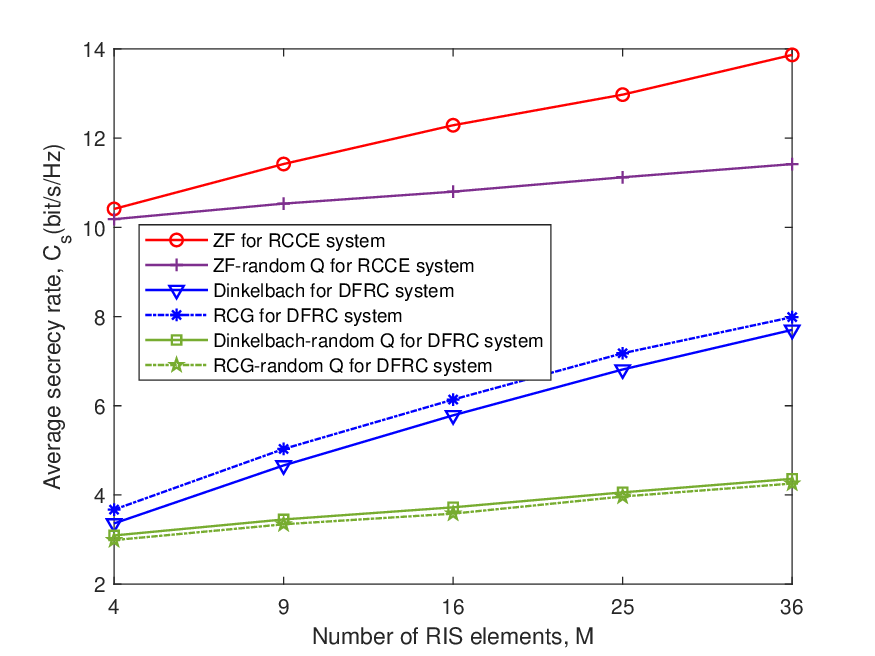}}
	\caption{Average secrecy rate $C_s$ v.s. number of RIS elements $M$.} \label{fig_M}
\end{figure}

In Fig. \ref{fig_M}, the secrecy rate versus the number of RIS elements is presented where the schemes with random $\bQ$ indicate that the other variables are optimized with a randomly chosen feasible $\bQ$. It can seen that in both systems, the schemes with joint optimization of active beamforming at the BS and passive beamforming of RIS achieve superior secrecy performance which highlights the effectiveness of the proposed algorithms and the significance of optimizing RIS phase shifts. Moreover, the secrecy rate monotonically increases with the number of RIS elements, since more RIS elements provide extra spatial degrees of freedom, enhancing the passive beamforming gain.

\subsection{RIS-aided Robust Secure Communication in DFRC System With Eve's Imperfect CSI}  
In this subsection, simulation results of the proposed algorithm in Section IV are provided. Unless otherwise stated, we set $P = 15$ dBm, $\gamma_{\mathrm{dB}} = \frac{\gamma_p}{\sigma_\mathrm{E}^2} = 0$ dB, %$\bar{\theta} = 30^{\circ}$, $\phi \!=\! 3^{\circ}$, 
and $\bar{\epsilon}_{\mathrm{G}} = \frac{\epsilon_{\mathrm{G}}}{\Vert\bar{\bG}\Vert_F}$. For simplicity, we set all the path losses as 1 and $\sigma_\mathrm{B}^2 = \sigma_\mathrm{E}^2 = 0$ dBm. The results in all figures are averaged over 100 randomly generated channels.

\begin{figure} [htbp]
	\centerline{\includegraphics[scale=0.62]{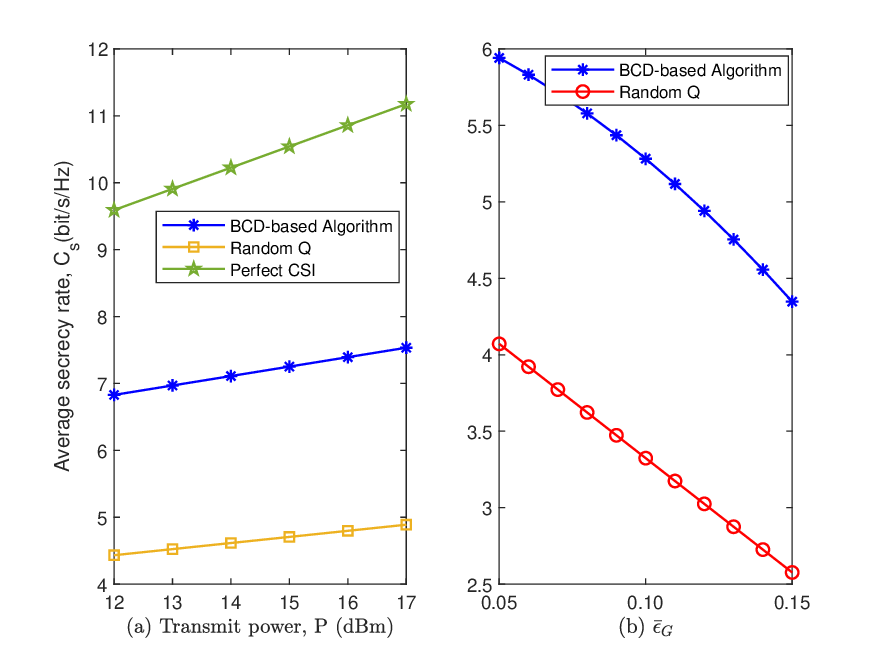}}
	\caption{Average secrecy rate $C_s$ v.s. (a) transmit power $P$ with $\phi = 2^{\circ}$, and (b) estimation uncertainty $\bar{\epsilon}_{\mathrm{G}}$ with $\phi = 3^{\circ}$.} \label{fig_epsilonIE}
\end{figure}

In Fig \ref{fig_epsilonIE} (a), the average secrecy rate versus the transmit power is plotted. {{We can observe that the average secrecy rate with target's imperfect CSI monotonically increases with the transmit power within the considered range while being upper-bounded by the perfect CSI counterpart}}\footnote{The results under perfect CSI are obtained by the Dinkelbach method-based AO algorithm described in Section III.}. 
%Furthermore, the proposed BCD-based algorithm enjoys a higher secrecy rate compared to the baseline scheme with a randomly chosen feasible $\bQ$. 
In Fig. \ref{fig_epsilonIE} (b), the average secrecy rate versus the estimation uncertainty $\bar{\epsilon}_{\mathrm{G}}$ is presented. It is evident that the average secrecy rate decreases with increasing $\bar{\epsilon}_{\mathrm{G}}$. Since with a larger $\bar{\epsilon}_{\mathrm{G}}$, the BS becomes less flexible when designing the beamforming. Additionally, over the entire considered range of $\bar{\epsilon}_{\mathrm{G}}$, the proposed BCD-based algorithm consistently outperforms the baseline scheme with a randomly chosen feasible $\bQ$, which verifies that the joint design of transmit beamforming and RIS's phase shifts can effectively achieve secure communication even if Eve's CSI is not perfectly known.

%\begin{figure} [htbp]
%	\centerline{\includegraphics[scale=0.62]{gamma_dB.eps}}
%	\caption{Average secrecy rate $C_s$ v.s. radar detection threshold $\gamma_{\mathrm{dB}}$.} \label{gamma_dB}
%\end{figure}
%Fig. \ref{gamma_dB} presents the variation of average secrecy rate with the radar detection threshold under imperfect/perfect CSI. It can be seen that the average secrecy rate in all cases decreases with $\gamma_{\mathrm{dB}}$.
% since a larger $\gamma_{\mathrm{dB}}$ means that more energy is required to illuminate the target, leading to an increasing eavesdropping information rate. As expected, the average secrecy rate with imperfect Eve's CSI is lower than that with perfect CSI, and for the cases of imperfect CSI, the average secrecy rate decreases with the uncertainty $\epsilon_{\mathrm{IE}}$. 
%Comparing the BCD-based algorithm and baseline scheme with random $\bQ$, we can find that the proposed algorithm can achieve a higher average secrecy rate over the entire considered range of $\gamma_{\mathrm{dB}}$ which manifests that RIS can help to strike a better balance between radar sensing and communication.  

\begin{figure} [htbp]
	\centerline{\includegraphics[scale=0.62]{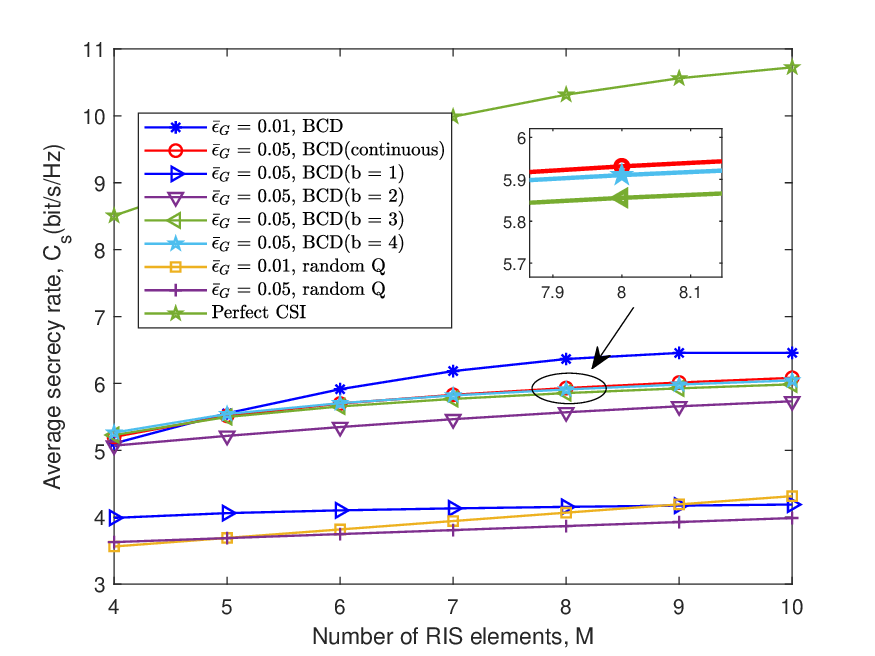}}
	\caption{Average secrecy rate $C_s$ v.s. number of RIS elements, $M$.} \label{M}
\end{figure}

{{Considering that RIS only has discrete phase shifts in practice, we further examine the impact of the discrete phase shift on the secrecy performance in Fig. \ref{M}. Similarly as done in \cite{8930608}, we quantify each optimal continuous phase shift to its nearest value in set $\mathcal{D}=\left\{0,\triangle\theta,\dots,\left(L-1\right)\triangle\theta\right\}, \triangle\theta=2\pi/L$, where discrete phase shifts are obtained by uniformly quantizing the interval $\left[0,2\pi\right)$ with $L=2^b$ being the total number of phase shifts and $b$ denoting the quantification bit number. 
From Fig. \ref{M}, it can easily seen that the highest average secrecy rate is achieved when the CSI is perfectly known and it grows almost linearly with the number of RIS elements. By contrast, with imperfect CSI, the average secrecy rate significantly decreases and exhibits slow growth with increasing $M$, since estimation error introduced by RIS restricts the effective exploration of beamforming gain. Moreover, comparing the curves of BCD algorithm with continuous and discrete phase shifts, it can be observed the secrecy rate loss when the quantification bit number $b$ is too small, e.g., $b=1$, whereas the secrecy rate under discrete phase shifts improves rapidly as $b$ increases, e.g., the secrecy rate loss is almost negligible for $b=4$. This implies that we can choose the appropriate quantization bit number to meet the system performance requirements in practical applications.}}
 
\section{Conclusions}
In this paper, the RIS-aided secure communication in integrated radar and communication system was investigated, where two systems were considered, namely, the RCCE system and the DFRC system. For both systems, the secrecy rate was maximized with maintaining radar sensing performance through joint design of transmit beamforming of the BS and the phase shift matrix of RIS. A ZF-based BCD algorithm was developed to solve the formulated problem for RCCE system. For the DFRC system, by exploiting the structures of objective function and optimization variables, two algorithms were proposed to tackle the secrecy rate maximization problem. Specially, a low-complexity RCG-based algorithm was developed. Moreover, the RIS-aided robust secure communication in DFRC system for practical case of target’s imperfect CSI was studied. The S-procedure was adopted to transform the semi-indefinite constraints incurred by CSI estimation errors and a BCD-based algorithm was proposed to address the formulated problem. Simulation results demonstrated that the RCCE system had a higher secrecy rate than DFRC system, and deploying RIS can enhance secure communication for both systems. When target’s CSI was not perfectly known, there was significant loss in secrecy rate, highlighting the importance of Eve’s accurate CSI for secure communication. Although introducing more reflecting elements can compensate for this loss, the secrecy rate with target's imperfect CSI increased slowly with the number of RIS elements. 

\small 
\bibliographystyle{IEEEtran}
\bibliography{reference}
\begin{IEEEbiography}[{\includegraphics[width=1in,height=1.25in,clip,keepaspectratio]{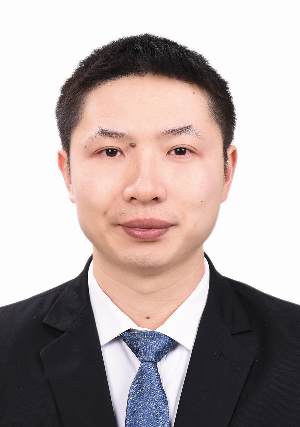}}]{Tong-Xing Zheng (Member, IEEE)} received the B.S. degree in information engineering and Ph.D. degree in information and communications engineering from Xi’an Jiaotong University, Xi’an, China, in 2010 and 2016, respectively. From 2017 to 2018, he was a Visiting Scholar with the School of Electrical Engineering and Telecommunications, University of New South Wales, Sydney, Australia. He is currently an Associate Professor with Xi’an Jiaotong University. His research interests include B5G/6G wireless networks, physical layer security, covert communications, reconfigurable intelligent surface, and integrated sensing and communications. He has published 3 book/chapters and 100+ papers in wireless communications. He was a recipient of the Excellent Doctoral Dissertation Award of Shaanxi Province in 2019, and the Best Paper Awards at The 2nd International Conference on Ubiquitous Communication (Ucom 2023) and The 12th International Conference on Information Systems and Computing Technology (ISCTech 2024). He was honored as an Exemplary Reviewer of \textsc{IEEE Transactions on Communications} in 2017, 2018, and 2021, respectively. He was a Leading Guest Editor of \textsc{Frontiers in Communications and Networks} for the Special Issue on Covert Communications for Next-Generation Wireless Networks in 2021 and a Guest Editor of \textsc{Wireless Communications and Mobile Computing} for the Special Issue on Physical Layer Security for Internet of Things in 2018. He served as an Associate Editor of \textsc{IET Electronic Letters} from 2020 to 2023. He served as a Session/Workshop Chair for the IEEE ICC 2022, IoTCIT 2024, ICCBD+AI 2024, and NCIC 2024. He has served as a TPC member for various IEEE sponsored conferences, including the GLOBECOM, ICC, WCNC, VTC, and PIMRC.
\end{IEEEbiography}

\begin{IEEEbiography}[{\includegraphics[width=1in,height=1.25in,clip,keepaspectratio]{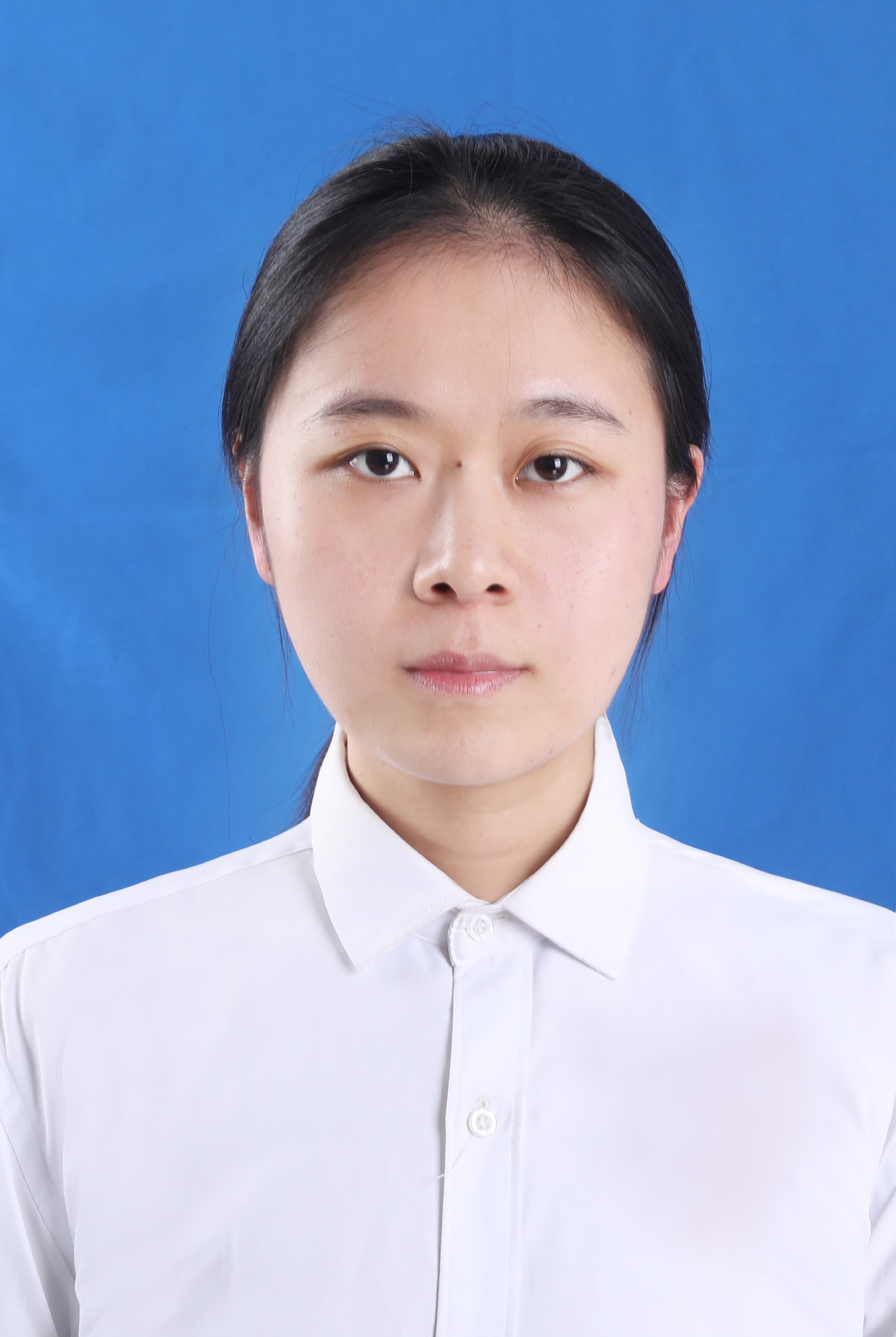}}]{Xin Chen} received the B.S. and M.S. degrees from the School of Information and Communications Engineering, Xi'an Jiaotong University, Xi’an, China, in 2021, and 2024, respectively, and she is currently with Purple Mountain Laboratories, Nanjing, China. Her research interests include wireless physical layer security, covert communication, reconfigurable intelligent surface, and integrated sensing and communications.
\end{IEEEbiography}

\begin{IEEEbiography}[{\includegraphics[width=1in,height=1.25in,clip,keepaspectratio]{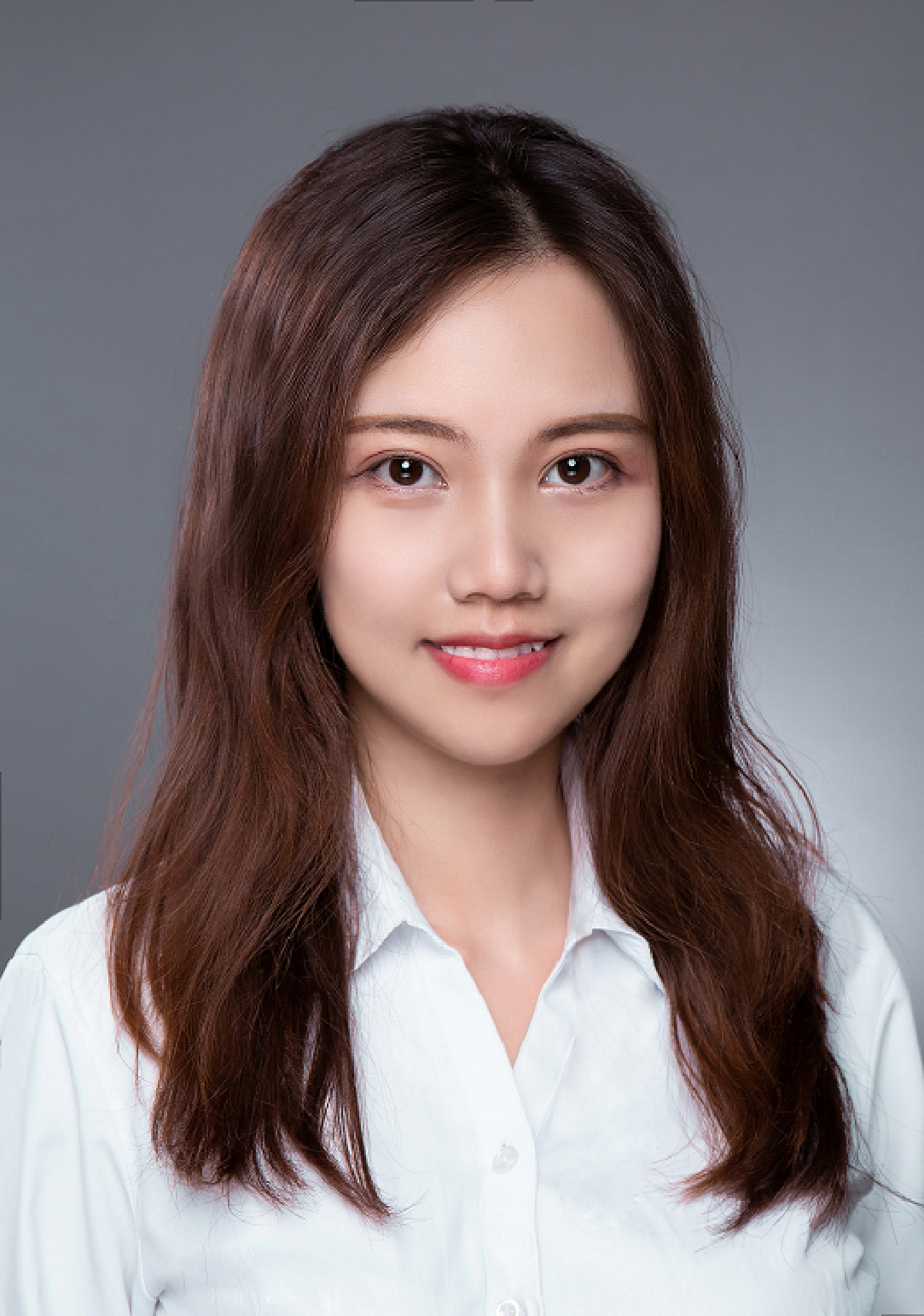}}]{Lan Lan (Member, IEEE)} received the B.S. degree in electronic engineering and the Ph.D. degree in signal and information processing from Xidian University, Xi'an, in 2015 and 2020, respectively. From July 2019 to July 2020, she was a Visiting Ph.D. Student with the University of Naples Federico II, Naples, Italy. She is currently an Associate Professor with the National Key Laboratory of Radar Signal Processing, Xidian University. Her research interests include frequency diverse array radar systems, MIMO radar signal processing, target detection, and ECCM.
	
Dr. Lan was elected as the Youth Elite Scientist Sponsorship Program by China Association for Science and Technology in 2022 and the XXXV-th URSI Young Scientists Award in 2023. She is the TPC Member and Session Chair of important conferences, including the ICASSP, IEEE Radar Conference, International Conference on Radar, and IEEE SAM. She is currently on the Editorial Board of \textsc{IEEE Transactions on Vehicular Technology} and \textsc{Digital Signal Processing}.
\end{IEEEbiography}

\begin{IEEEbiography}[{\includegraphics[width=1in,height=1.25in,clip,keepaspectratio]{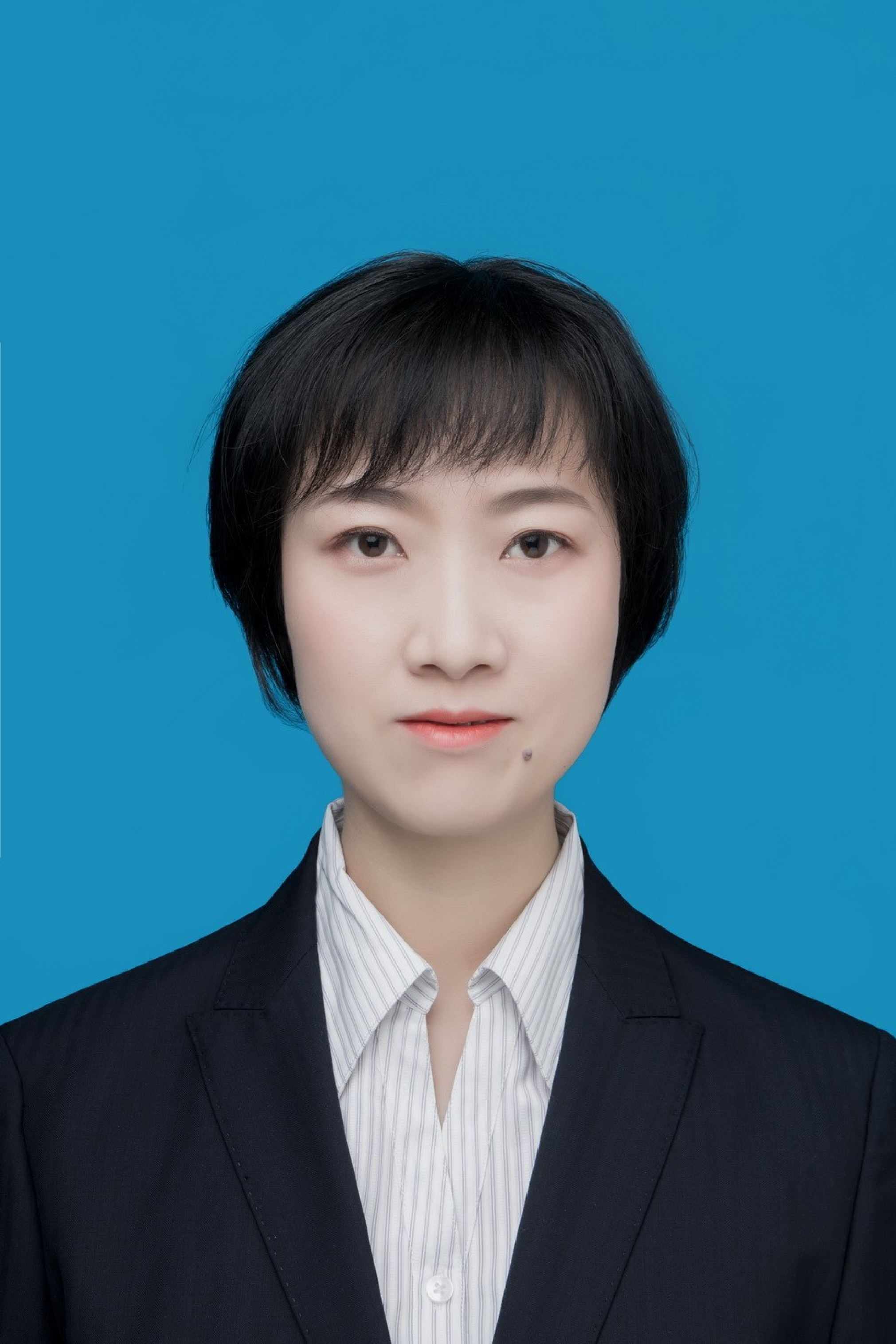}}]{Ying Ju (Member, IEEE)} received the B.S. and M.S. degrees from the School of Electronic Information Engineering, Tianjin University, Tianjin, China, in 2008 and 2010, respectively, and the Ph.D. degree from the School of Electronic and Information Engineering, Xi'an Jiaotong University, Xi'an, China, in 2018. From 2016 to 2017, she was a Visiting Scholar at the Department of Computer Science, University of California, Santa Barbara, USA. From 2010 to 2018, she was a Senior Engineer at the State Radio Monitoring Center, Xi'an, China. She is currently an Associate Professor with the Department of Telecommunications Engineering, Xidian University, Xi'an, China. Her research interests include physical layer security of wireless communications, millimeter wave communications, vehicular networks, and AI in wireless communication systems.
\end{IEEEbiography}

\begin{IEEEbiography}[{\includegraphics[width=1in,height=1.25in,clip,keepaspectratio]{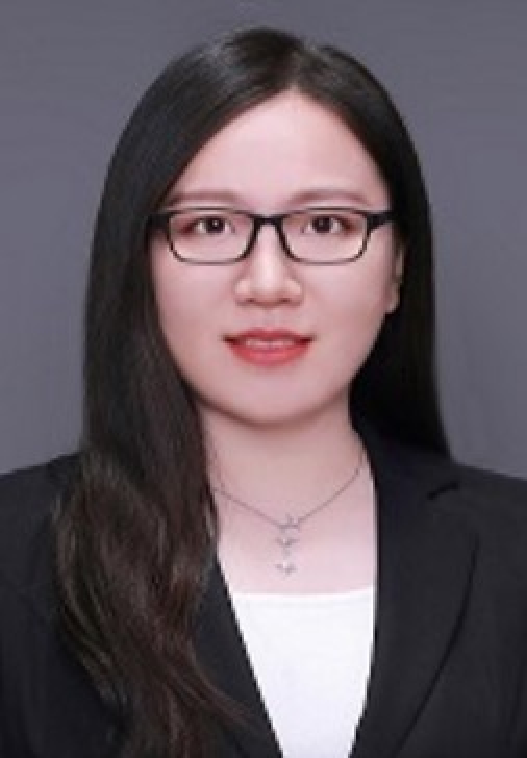}}]{Xiaoyan Hu (Member, IEEE)} received the Ph.D. degree in Electronic and Electrical Engineering from University College London (UCL), London, U.K., in 2020. From 2019 to 2021, she was a Research Fellow with the Department of Electronic and Electrical Engineering, UCL, U.K. She is currently an Associate Professor with the School of Information and Communications Engineering, Xi'an Jiaotong University, Xi'an, China.
Her research interests are in areas of 5G\&6G wireless communications, including edge computing, reconfigurable intelligent surface, UAV communications, ISAC, secure\&covert communications, and learning-based communications, etc. 
She has served as a Guest Editor for \textsc{Electronics} on Physical Layer Security and for China Communications Blue Ocean Forum on MAC and Networks. From 2020 to 2023, she served as the Assistant to the Editor-in-Chief of \textsc{IEEE Wireless Communications Letters}. She is currently serving as an Associate Editor for \textsc{IEEE Wireless Communications Letters}.
\end{IEEEbiography}

\begin{IEEEbiography}[{\includegraphics[width=1in,height=1.25in,clip,keepaspectratio]{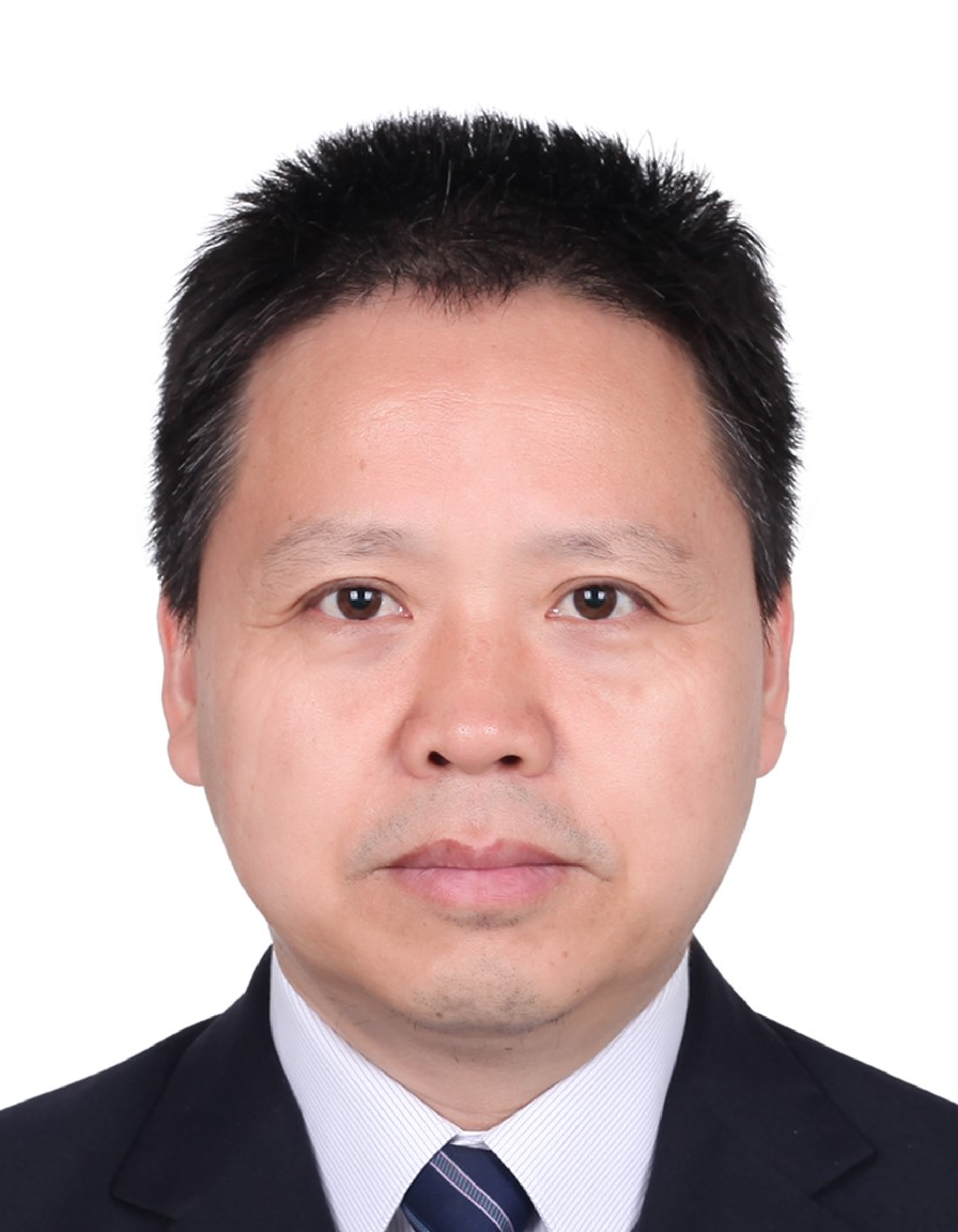}}]{Rongke Liu (Senior Member, IEEE)} received the B.S. and Ph.D. degrees from Beihang University, Beijing, China, in 1996 and 2002, respectively. He was a Visiting Professor with the Florida Institute of Technology, Melbourne, FL, USA, in 2006, The University of Tokyo, Tokyo, Japan, in 2015, and University of Edinburgh, Edinburgh, U.K., in 2018, respectively. He is currently a Full Professor with the School of Electronics and Information Engineering, Beihang University. He has authored or coauthored more than 200 papers in international conferences and journals. He has been granted more than 30 patents. His research interests include wireless communication (5G/B5G/6G) and satellite internet. He has attended many special programs, such as China Terrestrial Digital Broadcast Standard. He received the support of the New Century Excellent Talents Program from the Minister of Education, China.
\end{IEEEbiography}

\begin{IEEEbiography}[{\includegraphics[width=1in,height=1.25in,clip,keepaspectratio]{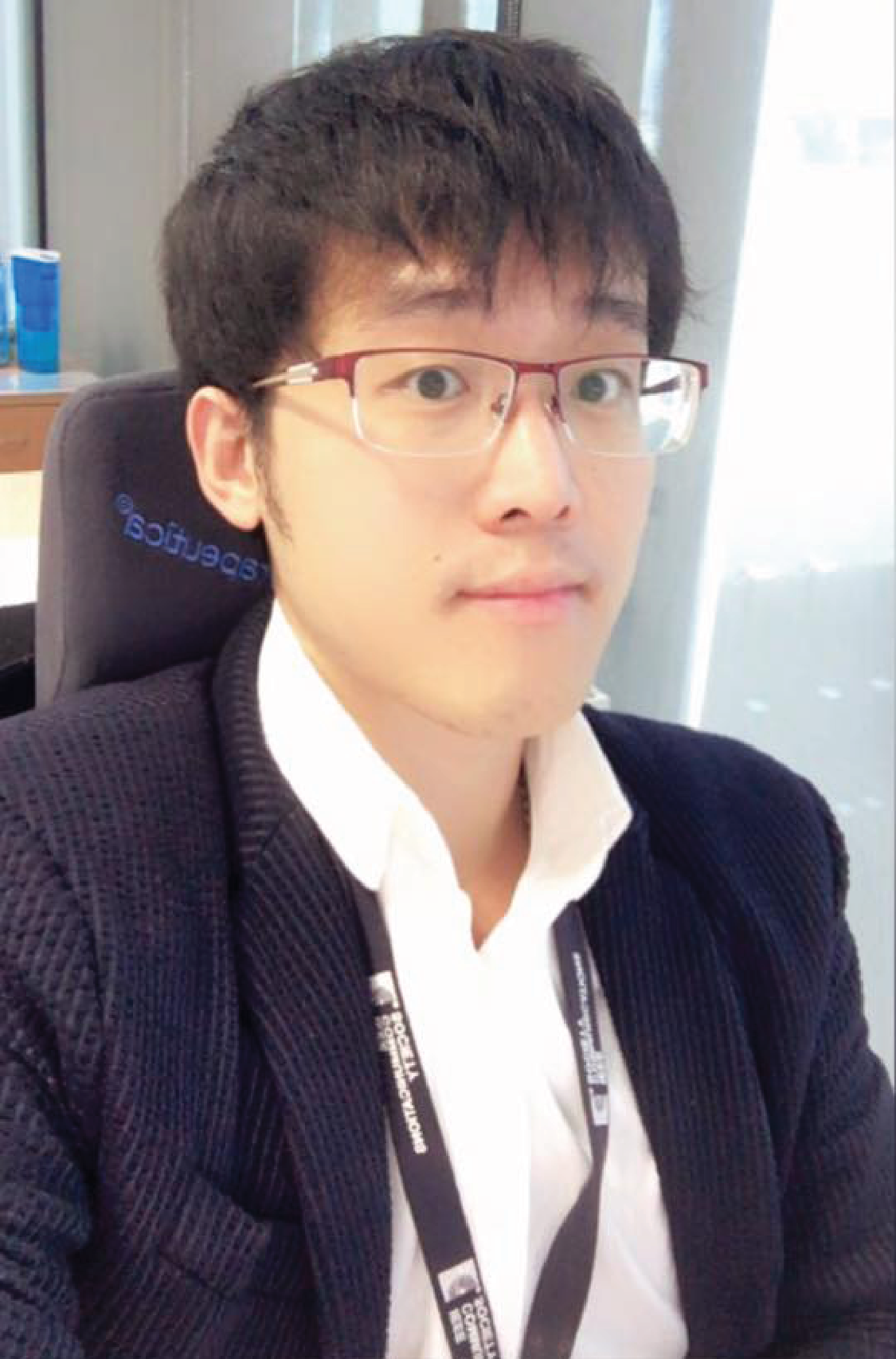}}]{Derrick Wing Kwan Ng (Fellow, IEEE)} received his bachelor's degree (with first-class Honors) and the Master of Philosophy degree in electronic engineering from The Hong Kong University of Science and Technology (HKUST), Hong Kong, in 2006 and 2008, respectively, and his Ph.D. degree from The University of British Columbia, Vancouver, BC, Canada, in November 2012. Following his Ph.D., he was a senior postdoctoral fellow at the Institute for Digital Communications, Friedrich-Alexander-University Erlangen-N\"urnberg (FAU), Germany. He is currently a Scientia Associate Professor with the University of New South Wales, Sydney, NSW, Australia. His research interests include global optimization, integrated sensing and communication (ISAC), physical layer security, IRS-assisted communication, UAV-assisted communication, wireless information and power transfer, and green (energy-efficient) wireless communications.
	
He has been recognized as a Highly Cited Researcher by Clarivate Analytics (Web of Science) since 2018. He was the recipient of the Australian Research Council (ARC) Discovery Early Career Researcher Award 2017, IEEE Communications Society Leonard G. Abraham Prize 2023, IEEE Communications Society Stephen O. Rice Prize 2022, Best Paper Awards at the WCSP 2020, 2021, IEEE TCGCC Best Journal Paper Award 2018, INISCOM 2018, IEEE International Conference on Communications (ICC) 2018, 2021, 2023, 2024,  IEEE International Conference on Computing, Networking and Communications (ICNC) 2016, IEEE Wireless Communications and Networking Conference (WCNC) 2012, IEEE Global Telecommunication Conference (Globecom) 2011, 2021, 2023 and IEEE Third International Conference on Communications and Networking in China 2008. From January 2012 to December 2019, he served as an Editorial Assistant to the Editor-in-Chief of the \textsc{IEEE Transactions on Communications}. He is also an Area Editor of the \textsc{IEEE Transactions on Communications} and an Associate Editor-in-Chief for the \textsc{IEEE Open Journal of the Communications Society}.
\end{IEEEbiography}

\begin{IEEEbiography}[{\includegraphics[width=1in,height=1.25in,clip,keepaspectratio]{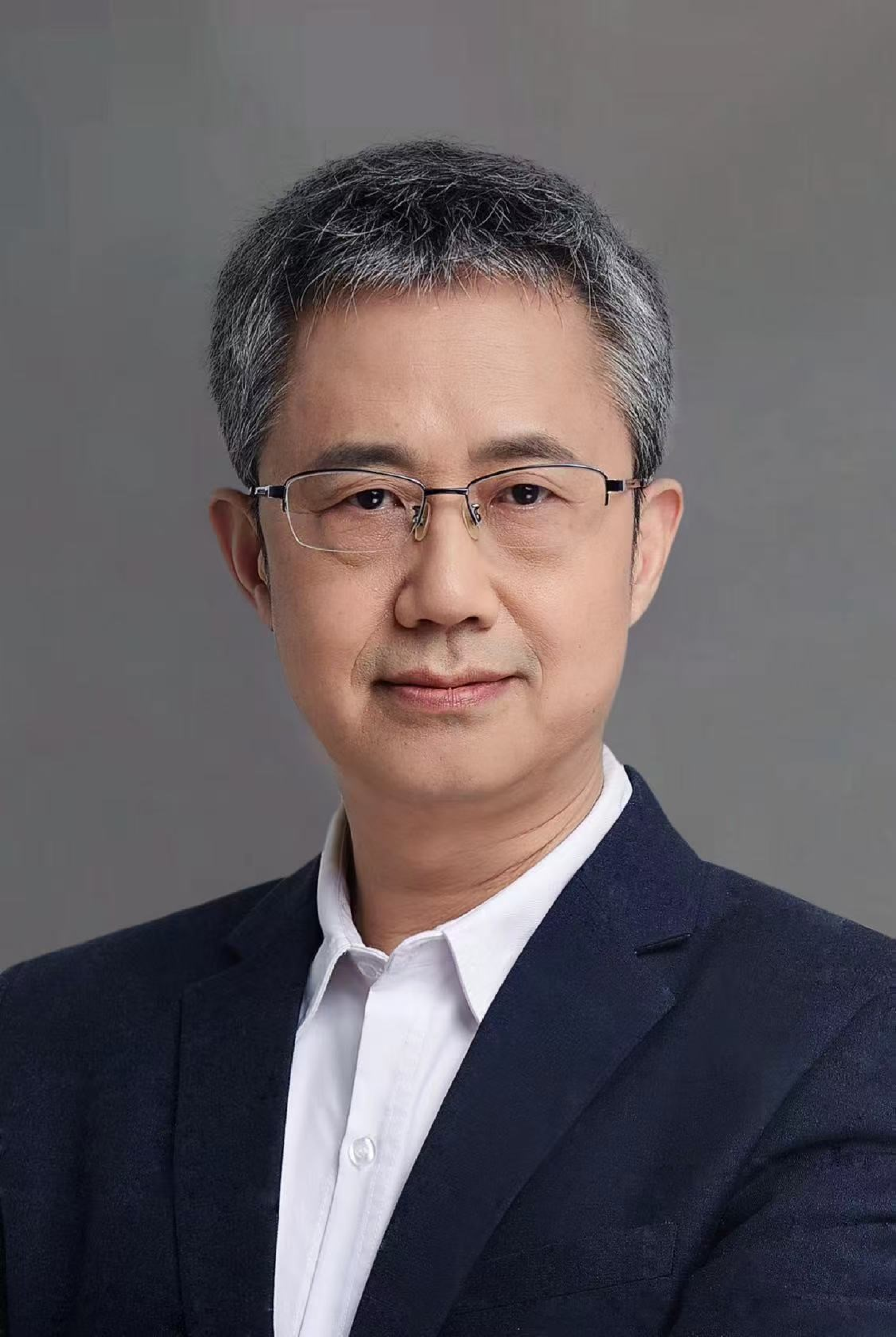}}]{Tiejun Cui (Fellow, IEEE)} received the B.Sc., M.Sc., and Ph.D. degrees in electrical engineering from Xidian University, Xi’an, China, in 1987, 1990, and 1993, respectively.
	
In 2001, he was a Cheung-Kong Professor with the Department of Radio Engineering, Southeast University, Nanjing, China. Currently he is the Chief Professor of Southeast University, and the Director of State Key Laboratory of Millimeter Waves. He is also the Founding Director of the Institute of Electromagnetic Space, Southeast University. His research interests include metamaterials and computational electromagnetics. He has published over 700 peer-review journal papers, which have been cited by more than 75000 times (H-Factor 132), and licensed over 160 patents. 
	
Dr. Cui is the Academician of Chinese Academy of Science, and IEEE Fellow. He served as Associate Editor of \textsc{IEEE Transactions on Geoscience and Remote Sensing}, and Guest Editors of \textsc{Science China-Information Sciences}, \textsc{Science Bulletin}, \textsc{IEEE Transactions on Microwave Theory and Techniques}, \textsc{IEEE Journal of Emerging Technologies in Circuits and System}, and \textsc{Applied Physics Letters}, \textsc{Engineering}, \textsc{Advanced Optical Materials}, and \textsc{Research}. Currently he is the Chief Editor of Metamaterial Short Books in Cambridge University Press, the Editor of \textsc{Materials Today Electronics}, and the Editorial Board Members or International Advisory Members of \textsc{National Science Review}, \textsc{eLight}, \textsc{PhotoniX}, \textsc{Small Structure}, \textsc{Physical Review Applied}, \textsc{Research}, \textsc{Advanced Optical Materials}, \textsc{Advanced Photonics Research}, and \textsc{Journal of Physics: Photonics}. He presented more than 100 Keynote and Plenary Talks in Academic Conferences, Symposiums, or Workshops.
\end{IEEEbiography}
\end{document}